\documentclass[10pt,prl,nofootinbib,preprint,superscriptaddress]{revtex4}

\usepackage{graphicx}
\usepackage{amsmath}
\usepackage{amsfonts}
\usepackage{amssymb}
\usepackage{dsfont}
\usepackage{dcolumn}
\usepackage{float}
\usepackage{bm}
\usepackage{pstricks}
\usepackage{color}
\usepackage{multirow}
\usepackage{booktabs}

\newcommand{\beqn}{\begin{eqnarray}}
\newcommand{\eeqn}{\end{eqnarray}}
\newcommand{\be}{\begin{equation}}
\newcommand{\ee}{\end{equation}}

\newcommand{\mathsym}[1]{{}}

\def\br{\left(\begin{array}{c}}
\def\er{\end{array}\right)}

\def\rmuu{\gamma^{\mu}}
\def\rmud{\gamma_{\mu}}
\def\PL{{1-\gamma_5\over 2}}
\def\PR{{1+\gamma_5\over 2}}
\def\sinW2{\sin^2\theta_W}
\def\AEM{\alpha_{EM}}
\def\mul{M_{\tilde{u} L}^2}
\def\mur{M_{\tilde{u} R}^2}
\def\mdl{M_{\tilde{d} L}^2}
\def\mdr{M_{\tilde{d} R}^2}
\def\mz2{M_{z}^2}
\def\c2b{\cos 2\beta}
\def\au{A_u}
\def\ad{A_d}
\def\cob{\cot \beta}
\def\v#1{v_#1}
\def\tb{\tan\beta}
\def\epem{$e^+e^-$}
\def\KK{$K^0$-$\overline{K^0}$}
\def\wi{\omega_i}
\def\xj{\chi_j}
\def\Wmu{W_\mu}
\def\Wnu{W_\nu}
\def\m#1{{\tilde m}_#1}
\def\mH{m_H}
\def\mw#1{{\tilde m}_{\omega #1}}
\def\mx#1{{\tilde m}_{\chi^{0}_#1}}
\def\mc#1{{\tilde m}_{\chi^{+}_#1}}
\def\mwi{{\tilde m}_{\omega i}}
\def\mxi{{\tilde m}_{\chi^{0}_i}}
\def\mci{{\tilde m}_{\chi^{+}_i}}

\def\ch{{\tilde\chi^{+}_1}}
\def\c2{{\tilde\chi^{+}_2}}

\def\tt{{\tilde\theta}}

\def\tp{{\tilde\phi}}

\def\mz{M_z}
\def\sw{\sin\theta_W}
\def\cw{\cos\theta_W}
\def\cb{\cos\beta}
\def\sb{\sin\beta}
\def\rwi{r_{\omega i}}
\def\rxj{r_{\chi j}}
\def\rfp{r_f'}
\def\Kik{K_{ik}}
\def\Fq2{F_{2}(q^2)}
\def\f{\({\cal F}\)}
\def\d1{{\f(\tilde c;\tilde s;\tilde W)+ \f(\tilde c;\tilde \mu;\tilde W)}}
\def\tw{\tan\theta_W}
\def\sec2w{sec^2\theta_W}
\def\lsim{\ ^<\llap{$_\sim$}\ }
\def\gsim{\ ^>\llap{$_\sim$}\ }
\def\r2{\sqrt 2}
\def\beq{\begin{equation}}
\def\eeq{\end{equation}}
\def\beqn{\begin{eqnarray}}
\def\eeqn{\end{eqnarray}}
\def\rmuu{\gamma^{\mu}}
\def\rmud{\gamma_{\mu}}
\def\PL{{1-\gamma_5\over 2}}
\def\PR{{1+\gamma_5\over 2}}
\def\sinW2{\sin^2\theta_W}
\def\AEM{\alpha_{EM}}
\def\mul{M_{\tilde{u} L}^2}
\def\mur{M_{\tilde{u} R}^2}
\def\mdl{M_{\tilde{d} L}^2}
\def\mdr{M_{\tilde{d} R}^2}
\def\mz2{M_{z}^2}
\def\c2b{\cos 2\beta}
\def\au{A_u}
\def\ad{A_d}
\def\cob{\cot \beta}
\def\v#1{v_#1}
\def\tb{\tan\beta}
\def\epem{$e^+e^-$}
\def\KK{$K^0$-$\bar{K^0}$}
\def\wi{\omega_i}
\def\xj{\chi_j}
\def\Wmu{W_\mu}
\def\Wnu{W_\nu}
\def\m#1{{\tilde m}_#1}
\def\mH{m_H}
\def\mw#1{{\tilde m}_{\omega #1}}
\def\mx#1{{\tilde m}_{\chi^{0}_#1}}
\def\mc#1{{\tilde m}_{\chi^{+}_#1}}
\def\mwi{{\tilde m}_{\omega i}}
\def\mxi{{\tilde m}_{\chi^{0}_i}}
\def\mci{{\tilde m}_{\chi^{+}_i}}
\def\mz{M_z}
\def\sw{\sin\theta_W}
\def\cw{\cos\theta_W}
\def\cb{\cos\beta}
\def\sb{\sin\beta}
\def\rwi{r_{\omega i}}
\def\rxj{r_{\chi j}}
\def\rfp{r_f'}
\def\Kik{K_{ik}}
\def\Fq2{F_{2}(q^2)}
\def\f{\({\cal F}\)}
\def\d1{{\f(\tilde c;\tilde s;\tilde W)+ \f(\tilde c;\tilde \mu;\tilde W)}}
\def\tw{\tan\theta_W}
\def\sec2w{sec^2\theta_W}
\def\ch{{\tilde\chi^{+}_1}}
\def\c2{{\tilde\chi^{+}_2}}

\def\tt{{\tilde\theta}}

\def\tp{{\tilde\phi}}

\def\mz{M_z}
\def\sw{\sin\theta_W}
\def\cw{\cos\theta_W}
\def\cb{\cos\beta}
\def\sb{\sin\beta}
\def\rwi{r_{\omega i}}
\def\rxj{r_{\chi j}}
\def\rfp{r_f'}
\def\Kik{K_{ik}}
\def\Fq2{F_{2}(q^2)}
\def\f{\({\cal F}\)}
\def\d1{{\f(\tilde c;\tilde s;\tilde W)+ \f(\tilde c;\tilde \mu;\tilde W)}}
\def\tw{\tan\theta_W}
\def\sec2w{sec^2\theta_W}

\def\rmuu{\gamma^{\mu}}
\def\rmud{\gamma_{\mu}}
\def\PL{{1-\gamma_5\over 2}}
\def\PR{{1+\gamma_5\over 2}}
\def\sinW2{\sin^2\theta_W}
\def\AEM{\alpha_{EM}}
\def\mul{M_{\tilde{u} L}^2}
\def\mur{M_{\tilde{u} R}^2}
\def\mdl{M_{\tilde{d} L}^2}
\def\mdr{M_{\tilde{d} R}^2}
\def\mz2{M_{z}^2}
\def\c2b{\cos 2\beta}
\def\au{A_u}
\def\ad{A_d}
\def\cob{\cot \beta}
\def\v#1{v_#1}
\def\tb{\tan\beta}
\def\epem{$e^+e^-$}
\def\KK{$K^0$-$\overline{K^0}$}
\def\wi{\omega_i}
\def\xj{\chi_j}
\def\Wmu{W_\mu}
\def\Wnu{W_\nu}
\def\m#1{{\tilde m}_#1}
\def\mH{m_H}
\def\mw#1{{\tilde m}_{\omega #1}}
\def\mx#1{{\tilde m}_{\chi^{0}_#1}}
\def\mc#1{{\tilde m}_{\chi^{+}_#1}}
\def\mwi{{\tilde m}_{\omega i}}
\def\mxi{{\tilde m}_{\chi^{0}_i}}
\def\mci{{\tilde m}_{\chi^{+}_i}}

\def\ch{{\tilde\chi^{+}_1}}
\def\c2{{\tilde\chi^{+}_2}}

\def\tt{{\tilde\theta}}

\def\tp{{\tilde\phi}}

\def\mz{M_z}
\def\sw{\sin\theta_W}
\def\cw{\cos\theta_W}
\def\cb{\cos\beta}
\def\sb{\sin\beta}
\def\rwi{r_{\omega i}}
\def\rxj{r_{\chi j}}
\def\rfp{r_f'}
\def\Kik{K_{ik}}
\def\Fq2{F_{2}(q^2)}
\def\f{\({\cal F}\)}
\def\d1{{\f(\tilde c;\tilde s;\tilde W)+ \f(\tilde c;\tilde \mu;\tilde W)}}
\def\tw{\tan\theta_W}
\def\sec2w{sec^2\theta_W}
\begin{document}
\baselineskip 18pt
\def\today{\ifcase\month\or
 January\or February\or March\or April\or May\or June\or
 July\or August\or September\or October\or November\or December\fi
 \space\number\day, \number\year}
\def\thebibliography#1{\section*{References\markboth
 {References}{References}}\list
 {[\arabic{enumi}]}{\settowidth\labelwidth{[#1]}
 \leftmargin\labelwidth
 \advance\leftmargin\labelsep
 \usecounter{enumi}}
 \def\newblock{\hskip .11em plus .33em minus .07em}
 \sloppy
 \sfcode`\.=1000\relax}
\let\endthebibliography=\endlist
\def\lsim{\ ^<\llap{$_\sim$}\ }
\def\gsim{\ ^>\llap{$_\sim$}\ }
\def\r2{\sqrt 2}
\def\beq{\begin{equation}}
\def\eeq{\end{equation}}
\def\beqn{\begin{eqnarray}}
\def\eeqn{\end{eqnarray}}
\def\rmuu{\gamma^{\mu}}
\def\rmud{\gamma_{\mu}}
\def\PL{{1-\gamma_5\over 2}}
\def\PR{{1+\gamma_5\over 2}}
\def\sinW2{\sin^2\theta_W}
\def\AEM{\alpha_{EM}}
\def\mul{M_{\tilde{u} L}^2}
\def\mur{M_{\tilde{u} R}^2}
\def\mdl{M_{\tilde{d} L}^2}
\def\mdr{M_{\tilde{d} R}^2}
\def\mz2{M_{z}^2}
\def\c2b{\cos 2\beta}
\def\au{A_u}
\def\ad{A_d}
\def\cob{\cot \beta}
\def\v#1{v_#1}
\def\tb{\tan\beta}
\def\epem{$e^+e^-$}
\def\KK{$K^0$-$\bar{K^0}$}
\def\wi{\omega_i}
\def\xj{\chi_j}
\def\Wmu{W_\mu}
\def\Wnu{W_\nu}
\def\m#1{{\tilde m}_#1}
\def\mH{m_H}
\def\mw#1{{\tilde m}_{\omega #1}}
\def\mx#1{{\tilde m}_{\chi^{0}_#1}}
\def\mc#1{{\tilde m}_{\chi^{+}_#1}}
\def\mwi{{\tilde m}_{\omega i}}
\def\mxi{{\tilde m}_{\chi^{0}_i}}
\def\mci{{\tilde m}_{\chi^{+}_i}}
\def\mz{M_z}
\def\sw{\sin\theta_W}
\def\cw{\cos\theta_W}
\def\cb{\cos\beta}
\def\sb{\sin\beta}
\def\rwi{r_{\omega i}}
\def\rxj{r_{\chi j}}
\def\rfp{r_f'}
\def\Kik{K_{ik}}
\def\Fq2{F_{2}(q^2)}
\def\f{\({\cal F}\)}
\def\d1{{\f(\tilde c;\tilde s;\tilde W)+ \f(\tilde c;\tilde \mu;\tilde W)}}
\def\tw{\tan\theta_W}
\def\sec2w{sec^2\theta_W}
\def\ch{{\tilde\chi^{+}_1}}
\def\c2{{\tilde\chi^{+}_2}}

\def\tt{{\tilde\theta}}

\def\tp{{\tilde\phi}}

\def\mz{M_z}
\def\sw{\sin\theta_W}
\def\cw{\cos\theta_W}
\def\cb{\cos\beta}
\def\sb{\sin\beta}
\def\rwi{r_{\omega i}}
\def\rxj{r_{\chi j}}
\def\rfp{r_f'}
\def\Kik{K_{ik}}
\def\Fq2{F_{2}(q^2)}
\def\f{\({\cal F}\)}
\def\d1{{\f(\tilde c;\tilde s;\tilde W)+ \f(\tilde c;\tilde \mu;\tilde W)}}
\def\tw{\tan\theta_W}
\def\sec2w{sec^2\theta_W}
\newcommand{\pn}[1]{{\color{red}{#1}}}

\begin{titlepage}

\begin{center}
{\large {\bf
Leptonic $g-2$ moments, CP phases and the Higgs boson mass constraint
}}\\
\vskip 0.5 true cm
\vspace{2cm}
\renewcommand{\thefootnote}
{\fnsymbol{footnote}}
Amin Aboubrahim$^a$\footnote{Email:a.aboubrahim@neu.edu},
 Tarek Ibrahim$^b$\footnote{Email:tibrahim@zewailcity.edu.eg}
 and Pran Nath$^a$\footnote{Email:p.nath@neu.edu}\\
 $^{a}$ Department of Physics, Northeastern University,
Boston, MA 02115-5000, USA\\
$^{b}$University of Science and Technology, Zewail City of Science and Technology,\\
 6th of October City, Giza 12588, Egypt\footnote{Permanent address:  Department of  Physics, Faculty of Science,
University of Alexandria, Alexandria 21511, Egypt}
\end{center}

\vskip 1.0 true cm

\centerline{\bf Abstract}
Higgs boson mass measurement at $\sim 125$ GeV points to a high scale for SUSY specifically
the scalar masses. If all the scalars are heavy, supersymmetric contribution to the leptonic $g-2$ moments will
be significantly reduced. On the other hand the Brookhaven experiment indicates a $\sim 3\sigma$
deviation from the standard model prediction. Here we analyze the leptonic $g-2$ moments in an extended
MSSM model with inclusion of a vector like leptonic generation which brings in new sources of CP violation.
{In this work we consider the contributions to the leptonic $g-2$ moments arising from the exchange of
charginos and neutralinos, sleptons and mirror sleptons, and from the exchange of $W$ and $Z$ bosons
and of leptons and mirror leptons.}
We focus specifically on the $g-2$ moments for the muon and the electron where sensitive measurements
 exist. Here it is shown that one can get consistency with the current data on $g-2$ under the Higgs boson mass constraint.
 Dependence of the moments on CP phases from the extended sector are analyzed and it is shown that they
 are sensitively dependent on the phases from the new sector. It is shown that the {corrections} to the leptonic
 moments arising from {the extended  MSSM}
 sector will be non-vanishing even if the SUSY scale extends into the PeV region. \\


\medskip
\end{titlepage}
\section{1. Introduction}\label{sec:intro}
The observation by ATLAS~\cite{Aad:2012tfa} and by CMS ~\cite{Chatrchyan:2012xdj}
 of the Higgs boson with a mass of $\sim 125$ GeV has put
very stringent constraints on low scale supersymmetry. Since the tree level mass of the Higgs boson
lies below $M_Z$,  a large loop correction from the supersymmetric sector is needed which
in turn implies a high scale for the weak scale supersymmetry and specifically for the  scalar masses.
A large SUSY scale also has direct implications for the $g_{\mu}-2$ of the muon. Thus the
current experimental result gives for the muon $g-2$~\cite{Beringer:1900zz}
\begin{align}
\Delta a_{\mu} =a_{\mu}^{\rm exp}- a_{\mu}^{\rm SM}=(26.2 \pm 8.5)\times 10^{-10}
\label{eq1}
\end{align}
 which is about  a three sigma deviation from the standard model prediction.  
 {Similarly for the electron the experimental determination of $g_e-2$ is
 very accurate and the uncertainty is rather small, i.e., one has~\cite{Giudice:2012ms}
\begin{align}
 \Delta a_e= a_e^{\rm exp}- a_e^{\rm SM}= -10.5 (8.1) \times  10^{-13}
 \label{eq2}
\end{align}
This result relies on a QED calculation up to four loops.
Thus along with Eq. (\ref{eq1}), Eq. (\ref{eq2})  also acts as a constraint on the standard model
extensions.
  Supersymmetric theories
 with  low  weak scale mass  can make corrections to $g_{\mu}-2$ which could be as large as the standard
 model electroweak  corrections and even larger and have strong CP phase dependence
 \cite{Ibrahim:1999hh,Ibrahim:1999aj,Ibrahim:2001ym}
 (for early work see \cite{Yuan:1984ww}).
 These arise largely from the chargino and sneutrino
 exchange diagram with the neutralino and smuon exchange diagram making a relatively small
 contribution. However, if the scalar masses are large, the supersymmetric exchange contributions
 will be small due to the largeness of the sneutrino and the  smuon {masses}. }


In this work we give an analysis of the $g-2$ for the muon and for the electron in
an extended MSSM model with a vector like leptonic generation.  {We note that vector like multiplets
are anomaly free and they appear} in a variety of settings which include grand unified models,
strings and D brane models~\cite{Ibrahim:2008gg,vectorlike,Babu:2008ge,Liu:2009cc,Martin:2009bg}.
Further, it is known that $g-2$ has a sharp dependence on CP phases~\cite{Ibrahim:1999hh,Ibrahim:1999aj,Ibrahim:2001ym}.
For this reason we investigate also the dependence of the muon and the electron $g-2$ on the CP phases
in the extended MSSM model.  Here we are particularly interested in the dependence on the CP phases
that arise from the new sector involving vector like leptons. We note that the CP phases are
 {constrained}
 in this case by the electric dipole moment of the electron which currently has the
value $|d_e|  < 8.7\times 10^{-29}$ $e$cm~\cite{Baron:2013eja}
 while the upper limit on the muon EDM is
$|d_\mu| < 1.9\times 10^{-19}$$e$cm~\cite{Beringer:1900zz} and is rather weak.
As discussed in several works even with large phases  the EDMs can be suppressed
 either by mass suppression~\cite{Nath:1991dn,Kizukuri:1992nj} or via the cancellation mechanism~\cite{Ibrahim:1998je,Ibrahim:1997gj,Falk:1998pu,Ibrahim:1998je,Brhlik:1998zn,Ibrahim:1999af}.
 Several analyses of the vector like extensions of MSSM already exist in the literature
\cite{Ibrahim:2010va,Ibrahim:2010hv,Ibrahim:2011im,Ibrahim:2012ds,Aboubrahim:2013gfa,Dermisek:2013gta,Nickel:2015dna,Aboubrahim:2015zpa,Ibrahim:2015hva,Ibrahim:2016rcb}.

The outline of the rest of the paper is as follows: In section 2  we give an analytical computation
for the contribution of the  vectorlike lepton generation to $g-2$ of the muon and of the electron.
 In section (3) we give a numerical analysis of the  contributions arising from MSSM and from
 the extended MSSM with  a vector like leptonic generation.
 Conclusions are given in section 4. Details of the extended MSSM model with a vector like leptonic generation are given in the Appendix. {The explanation of the muon anomaly with vector like leptons was considered
 previously in ~\cite{Dermisek:2013gta} within a non-supersymmetric framework. Our analysis is within a
 supersymmetric framework where we carry out a simultaneous fit to both the muon as well as the electron
 anomaly. Further, we explore the implications of the CP phases arising from the new sector.}

\section{2. Analysis of $g_\mu-2$  and $g_e-2$ with exchange of vector like leptons\label{secgmue}  }

The extended MSSM with a vector like leptonic generation is discussed in detail in the Appendix. Using the
formalism described there we compute the contribution to the  anomalous magnetic moment of
a charged lepton $\ell_\alpha$. We discuss now in detail the various contributions. The contribution
arising from the exchange of the charginos,  sneutrinos and mirror sneutrinos  as shown in the left diagram in Fig. 1 is
given by

\begin{align}
a_{\alpha}^{\chi^{+}}&=-\sum_{i=1}^{2}\sum_{j=1}^{10}\frac{m_{\tau_{\alpha}}}{16\pi^{2}m_{\chi_{i}^{-}}}\text{Re}(C^{L}_{\alpha ij}C^{R*}_{\alpha ij})
F_{3}\left(\frac{m^{2}_{\tilde{\nu}_{j}}}{m^{2}_{\chi^{-}_{i}}}\right) \nonumber \\
&+\sum_{i=1}^{2}\sum_{j=1}^{10}\frac{m^{2}_{\tau_{\alpha}}}{96\pi^{2}m^{2}_{\chi_{i}^{-}}}\left[|C^{L}_{\alpha ij}|^{2}+|C^{R}_{\alpha ij}|^{2}\right]
F_{4}\left(\frac{m^{2}_{\tilde{\nu}_{j}}}{m^{2}_{\chi^{-}_{i}}}\right),
\label{3}
\end{align}
{where $m_{\chi_{i}^{-}}$ is the mass of chargino $\chi_{i}^{-}$ and $m_{\tilde{\nu}_{j}}$  is the mass of 
sneutrino ${\tilde{\nu}_{j}}$} and where the form factors $F_3$ and $F_4$ are given by

\begin{align}
F_{3}(x)&=\frac{1}{(x-1)^{3}}\left[3x^{2}-4x+1-2x^{2}\ln x \right]
\end{align}
and
\begin{align}
F_{4}(x)&=\frac{1}{(x-1)^{4}}\left[2x^{3}+3x^{2}-6x+1-6x^{2}\ln x \right]
\end{align}

The couplings appearing in Eq. (\ref{3}) are given by 
\begin{align}
\begin{split}
C_{\alpha ij}^{L}=&g(-\kappa_{\tau}U^{*}_{i2}D^{\tau*}_{R1\alpha} \tilde{D}^{\nu}_{1j} -\kappa_{\mu}U^{*}_{i2}D^{\tau*}_{R3\alpha}\tilde{D}^{\nu}_{5j}-
\kappa_{e}U^{*}_{i2}D^{\tau*}_{R4\alpha}\tilde{D}^{\nu}_{7j}\\
&-\kappa_{4\ell}U^{*}_{i2}D^{\tau*}_{R5\alpha}\tilde{D}^{\nu}_{9j}
+U^{*}_{i1}D^{\tau*}_{R2\alpha}\tilde{D}^{\nu}_{4j}-
\kappa_{N}U^{*}_{i2}D^{\tau*}_{R2\alpha}\tilde{D}^{\nu}_{2j})
\end{split} \\~\nonumber\\
\begin{split}
C_{\alpha ij}^{R}=&g(-\kappa_{\nu_{\tau}}V_{i2}D^{\tau*}_{L1\alpha}\tilde{D}^{\nu}_{3j}-\kappa_{\nu_{\mu}}V_{i2}D^{\tau*}_{L3\alpha}\tilde{D}^{\nu}_{6j}-
\kappa_{\nu_{e}}V_{i2}D^{\tau*}_{L4\alpha}\tilde{D}^{\nu}_{8j}+V_{i1}D^{\tau*}_{L1\alpha}\tilde{D}^{\nu}_{1j}+V_{i1}D^{\tau*}_{L3\alpha}\tilde{D}^{\nu}_{5j}\\
&-\kappa_{\nu_{4}}V_{i2}D^{\tau*}_{L5\alpha}\tilde{D}^{\nu}_{10j}
+V_{i1}D^{\tau*}_{L4\alpha}\tilde{D}^{\nu}_{7j}-\kappa_{E}V_{i2}D^{\tau*}_{L2\alpha}\tilde{D}^{\nu}_{4j}),
\end{split}
\end{align}
{where $D_{L,R}^\tau$ and $\tilde D^\nu$ are the charged lepton and sneutrino diagonalizing matrices and 
 are defined by Eq. (\ref{Dtau}) and Eq.(\ref{Dtildenu}) and $U$ and $V$ are the matrices that diagonalize the
 chargino mass matrix $M_C$  so that~\cite{Ibrahim:2007fb} 
 \beq
  U^* M_C V^{-1} = diag(m_{\chi_1^{\pm}} m_{\chi_2^{\pm}})\,.
 \eeq 
Further, 
}

\begin{align}
(\kappa_{N},\kappa_{\tau},\kappa_{\mu},\kappa_{e},\kappa_{4\ell})&=\frac{(m_{N},m_{\tau},m_{\mu},m_{e},m_{4\ell})}{\sqrt{2}m_{W}\cos\beta} , \\~\nonumber\\
(\kappa_{E},\kappa_{\nu_{\tau}},\kappa_{\nu_{\mu}},\kappa_{\nu_{e}},\kappa_{\nu_{4}})&=\frac{(m_{E},m_{\nu_{\tau}},m_{\nu_{\mu}},m_{\nu_{e}},m_{\nu_{4}})}{\sqrt{2}m_{W}\sin\beta} .
\end{align}
{where $m_W$ is the mass of the $W$ boson and $\tan\beta= <H_2^2>/<H_1^1>$ where $H_1, H_2$ are the
two Higgs doublets of MSSM.}

\begin{figure}[t]
\begin{center}
{\rotatebox{0}{\resizebox*{10cm}{!}{\includegraphics{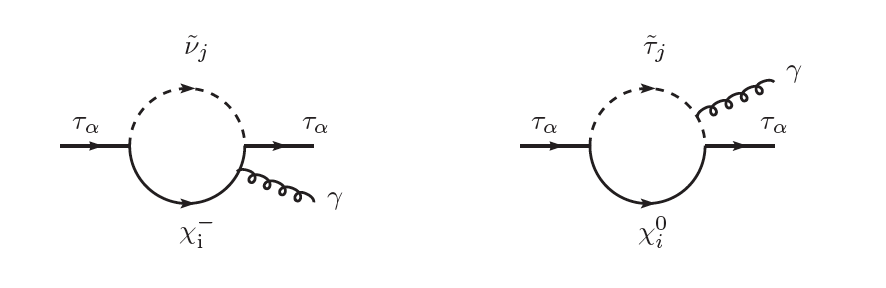}}\hglue5mm}}
\caption{The diagrams that contribute to the leptonic ($\tau_{\alpha}$)
magnetic dipole moment via
  exchange of charginos ($\chi_i^{-}$), sneutrinos and mirror sneutrinos  ($\tilde \nu_j$)   (left diagram) inside the loop and from the exchange
  of neutralinos ($\chi_i^0$),   sleptons   and mirror sleptons ($\tilde \tau_j$) (right diagram) inside the loop.} \label{fig1}
\end{center}
\end{figure}

\begin{figure}[t]
\begin{center}
{\rotatebox{0}{\resizebox*{10cm}{!}{\includegraphics{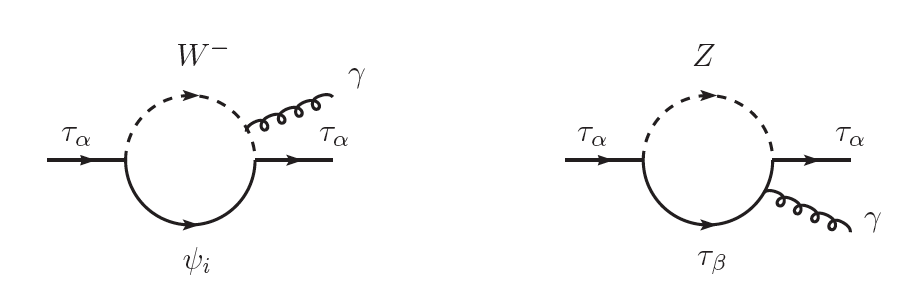}}\hglue5mm}}
\caption{ The W loop  (the left diagram) involving the exchange of sequential and vectorlike neutrinos $\psi_i$
and the Z loop (the right diagram) involving the exchange of sequential and vectorlike charged leptons $\tau_{\beta}$
that contribute to the magnetic dipole moment of the charged lepton $\tau_{\alpha}$.}
\label{fig2}
\end{center}
\end{figure}

The contribution  arising from the exchange of neutralinos, charged sleptons and charged mirror sleptons
as shown in the right diagram in Fig. 1 is given by

\begin{align}
a_{\alpha}^{\chi^{0}}&=\sum_{i=1}^{4}\sum_{j=1}^{10}\frac{m_{\tau_{\alpha}}}{16\pi^{2}m_{\chi_{i}^{0}}}\text{Re}(C^{'L}_{\alpha ij}C^{'R*}_{\alpha ij})
F_{1}\left(\frac{m^{2}_{\tilde{\tau}_{j}}}{m^{2}_{\chi^{0}_{i}}}\right) \nonumber \\
&+\sum_{i=1}^{4}\sum_{j=1}^{10}\frac{m^{2}_{\tau_{\alpha}}}{96\pi^{2}m^{2}_{\chi_{i}^{0}}}\left[|C^{'L}_{\alpha ij}|^{2}+|C^{'R}_{\alpha ij}|^{2}\right]
F_{2}\left(\frac{m^{2}_{\tilde{\tau}_{j}}}{m^{2}_{\chi^{0}_{i}}}\right),
\label{10}
\end{align}
where the form factors are

\begin{align}
F_{1}(x)&=\frac{1}{(x-1)^{3}}\left[1-x^{2}+2x\ln x \right]
\end{align}
and
\begin{align}
F_{2}(x)&=\frac{1}{(x-1)^{4}}\left[-x^{3}+6x^{2}-3x-2-6x\ln x \right]
\end{align}
 The couplings  that enter in Eq. \ref{10} are given by

\begin{align}
C_{\alpha ij}^{'L}=&\sqrt{2}(\alpha_{\tau i}D^{\tau *}_{R1\alpha}\tilde{D}^{\tau}_{1j}-\delta_{E i}D^{\tau *}_{R2\alpha}\tilde{D}^{\tau}_{2j}-
\gamma_{\tau i}D^{\tau *}_{R1\alpha}\tilde{D}^{\tau}_{3j}+\beta_{E i}D^{\tau *}_{R2\alpha}\tilde{D}^{\tau}_{4j}
+\alpha_{\mu i}D^{\tau *}_{R3\alpha}\tilde{D}^{\tau}_{5j}-\gamma_{\mu i}D^{\tau *}_{R3\alpha}\tilde{D}^{\tau}_{6j} \nonumber\\
&+\alpha_{e i}D^{\tau *}_{R4\alpha}\tilde{D}^{\tau}_{7j}-\gamma_{e i}D^{\tau *}_{R4\alpha}\tilde{D}^{\tau}_{8j}
+\alpha_{4\ell i}D^{\tau *}_{R5\alpha}\tilde{D}^{\tau}_{9j}-\gamma_{4\ell i}D^{\tau *}_{R5\alpha}\tilde{D}^{\tau}_{10j}
)
\label{CprimeL}
\end{align}
\begin{align}
C_{\alpha ij}^{'R}=&\sqrt{2}(\beta_{\tau i}D^{\tau *}_{L1\alpha}\tilde{D}^{\tau}_{1j}-\gamma_{E i}D^{\tau *}_{L2\alpha}\tilde{D}^{\tau}_{2j}-
\delta_{\tau i}D^{\tau *}_{L1\alpha}\tilde{D}^{\tau}_{3j}+\alpha_{E i}D^{\tau *}_{L2\alpha}\tilde{D}^{\tau}_{4j}
+\beta_{\mu i}D^{\tau *}_{L3\alpha}\tilde{D}^{\tau}_{5j}-\delta_{\mu i}D^{\tau *}_{L3\alpha}\tilde{D}^{\tau}_{6j}      \nonumber\\
&+\beta_{e i}D^{\tau *}_{L4\alpha}\tilde{D}^{\tau}_{7j}-\delta_{e i}D^{\tau *}_{L4\alpha}\tilde{D}^{\tau}_{8j}
+\beta_{4\ell i}D^{\tau *}_{L5\alpha}\tilde{D}^{\tau}_{9j}-\delta_{4\ell i}D^{\tau *}_{L5\alpha}\tilde{D}^{\tau}_{10j}
),
\label{CprimeR}
\end{align}
where

\begin{align}
\alpha_{E i}&=\frac{gm_{E}X^{*}_{4i}}{2m_{W}\sin\beta} \ ;  && \beta_{E i}=eX'_{1i}+\frac{g}{\cos\theta_{W}}X'_{2i}\left(\frac{1}{2}-\sin^{2}\theta_{W}\right) \\
\gamma_{E i}&=eX^{'*}_{1i}-\frac{g\sin^{2}\theta_{W}}{\cos\theta_{W}}X^{'*}_{2i} \  ;  && \delta_{E i}=-\frac{gm_{E}X_{4i}}{2m_{W}\sin\beta}
\end{align}

and
\begin{align}
\alpha_{\tau i}&=\frac{gm_{\tau}X_{3i}}{2m_{W}\cos\beta} \ ;  && \alpha_{\mu i}=\frac{gm_{\mu}X_{3i}}{2m_{W}\cos\beta} \ ; && \alpha_{e i}=\frac{gm_{e}X_{3i}}{2m_{W}\cos\beta}
 ; && \alpha_{4\ell i}=\frac{gm_{4\ell}X_{3i}}{2m_{W}\cos\beta}
 \\
\delta_{\tau i}&=-\frac{gm_{\tau}X^{*}_{3i}}{2m_{W}\cos\beta} \ ; && \delta_{\mu i}=-\frac{gm_{\mu}X^{*}_{3i}}{2m_{W}\cos\beta} \ ; && \delta_{e i}=-\frac{gm_{e}X^{*}_{3i}}{2m_{W}\cos\beta}
 ; && \delta_{4\ell i}=-\frac{gm_{4\ell}X^{*}_{3i}}{2m_{W}\cos\beta}
\end{align}
{and where }

\begin{align}
\beta_{\tau i}=\beta_{\mu i}=\beta_{e i}=\beta_{4\ell i}&=-eX^{'*}_{1i}+\frac{g}{\cos\theta_{W}}X^{'*}_{2i}\left(-\frac{1}{2}+\sin^{2}\theta_{W}\right)  \\
\gamma_{\tau i}=\gamma_{\mu i}=\gamma_{e i}=\gamma_{4\ell i}&=-eX'_{1i}+\frac{g\sin^{2}\theta_{W}}{\cos\theta_{W}}X'_{2i}
\end{align}
Here $X'$ are defined by

\begin{align}
X'_{1i}&=X_{1i}\cos\theta_{W}+X_{2i}\sin\theta_{W}  \\
X'_{2i}&=-X_ {1i}\sin\theta_{W}+X_{2i}\cos\theta_{W}
\end{align}
where $X$ diagonalizes the neutralino mass matrix, i.e.,

\beqn
X^{T}M_{\chi^{0}}X=\text{diag}(m_{\chi^{0}_{1}},m_{\chi^{0}_{2}},m_{\chi^{0}_{3}},m_{\chi^{0}_{4}}).
\eeqn
{Further, $\tilde D^\tau$ that enter in Eqs. (\ref{CprimeL}) and (\ref{CprimeR}) 
is a matrix which diagonalizes the charged slepton mass squared matrix 
and is defined in Eq. (\ref{Dtildetau}.). }

Next we compute the contribution from the exchange of the $W$ and $Z$ bosons.
Thus the exchange of the $W$ and the exchange of neutrinos and mirror neutrinos as shown
in the left diagram of Fig. 2 gives

\beqn
a^W_{\tau_{\alpha}} = \frac{m^2_{\tau_{\alpha}}}{16 \pi^2 m^2_W} \sum_{i=1}^{5} [|C^W_{Li\alpha}|^2
+|C^W_{Ri\alpha}|^2] F_W \left(\frac{m^2_{\psi_i}}{m^2_W}\right) + \frac{m_{\psi_i}}{m_{\tau_{\alpha}}} \text{Re}(C^W_{L i \alpha}C^{W*}_{R i \alpha}) G_W \left(\frac{m^2_{\psi_i}}{m^2_W}\right),
\label{24}
\eeqn
where the form factors are given by
\begin{align}
F_{W}(x)&=\frac{1}{6(x-1)^{4}}\left[4 x^4- 49x^{3}+18 x^3 \ln x+78x^{2}-43 x +10 \right]
\end{align}
and
\begin{align}
G_{W}(x)&=\frac{1}{(x-1)^{3}}\left[4 -15 x+12 x^2 - x^3-6 x^2 \ln x \right]
\end{align}

The couplings  that enter in Eq. (\ref{24}) are given by
\beqn
C_{L_{i\alpha}}^W= \frac{g}{\sqrt{2}} [D^{\nu*}_{L1i}D^{\tau}_{L1\alpha}+
D^{\nu*}_{L3i}D^{\tau}_{L3\alpha}+D^{\nu*}_{L4i}D^{\tau}_{L4\alpha}
+D^{\nu*}_{L5i}D^{\tau}_{L5\alpha}]  \\
C_{R_{i\alpha}}^W= \frac{g}{\sqrt{2}}[D^{\nu*}_{R2i}D^{\tau}_{R2\alpha}]
\eeqn
{Here $D_{L,R}^\nu$ are matrices of  a bi-unitary transformation that diagonalizes the neutrino mass matrix 
and are defined in Eq. (\ref{Dnu}).}

Finally the exchange of the $Z$ and the exchange of leptons  and mirror leptons  as shown
in the right diagram of Fig. 2 gives

\beqn
a^Z_{\tau_{\alpha}} = \frac{m^2_{\tau_{\alpha}}}{32 \pi^2 m^2_Z} \sum_{\beta=1}^{5} [|C^Z_{L \beta \alpha}|^2
+|C^Z_{R\beta \alpha}|^2] F_Z \left(\frac{m^2_{\tau_{\beta}}}{m^2_Z}\right) + \frac{m_{\tau_{\beta}}}{m_{\tau_{\alpha}}} \text{Re}(C^Z_{L \beta \alpha}C^{Z*}_{R \beta \alpha}) G_Z \left(\frac{m^2_{\tau_{\beta}}}{m^2_Z}\right),
\label{29}
\eeqn
where
\begin{align}
F_{Z}(x)&=\frac{1}{3(x-1)^{4}}\left[-5 x^4+14x^{3}-39 x^2+18 x^2 \ln x+38 x -8 \right]
\end{align}
and
\begin{align}
G_{Z}(x)&=\frac{2}{(x-1)^{3}}\left[x^3 + 3 x-6 x \ln x-4 \right],
\end{align}
{and $m_Z$  is the $Z$ boson mass.}
 The couplings  that enter in Eq. (\ref{29}) are given by

\beqn
C_{L_{\alpha \beta}}^Z=\frac{g}{\cos\theta_{W}} [x(D_{L\alpha 1}^{\tau\dag}D_{L1\beta}^{\tau}+D_{L\alpha 2}^{\tau\dag}D_{L2\beta}^{\tau}+D_{L\alpha 3}^{\tau\dag}D_{L3\beta}^{\tau}+D_{L\alpha 4}^{\tau\dag}D_{L4\beta}^{\tau}
+D_{L\alpha 5}^{\tau\dag}D_{L5\beta}^{\tau}
)\nonumber\\
-\frac{1}{2}(D_{L\alpha 1}^{\tau\dag}D_{L1\beta}^{\tau}+D_{L\alpha 3}^{\tau\dag}D_{L3\beta}^{\tau}+D_{L\alpha 4}^{\tau\dag}D_{L4\beta}^{\tau}
+D_{L\alpha 5}^{\tau\dag}D_{L5\beta}^{\tau}
)]
\eeqn
and
\beqn
C_{R_{\alpha \beta}}^Z=\frac{g}{\cos\theta_{W}} [x(D_{R\alpha 1}^{\tau\dag}D_{R1\beta}^{\tau}+D_{R\alpha 2}^{\tau\dag}D_{R2\beta}^{\tau}+D_{R\alpha 3}^{\tau\dag}D_{R3\beta}^{\tau}+D_{R\alpha 4}^{\tau\dag}D_{R4\beta}^{\tau}
+D_{R\alpha 5}^{\tau\dag}D_{R5\beta}^{\tau}
)\nonumber\\
-\frac{1}{2}(D_{R\alpha 2}^{\tau\dag}
D_{R 2\beta }^{\tau}
 )]\,.
\eeqn


\section{3. Estimates of $\Delta a_{\mu}$  and $\Delta a_e$  \label{secnum} }

We begin by discussing the prediction for $\Delta a_\mu$ and $\Delta a_e$ for  MSSM when the scalar masses are large lying
in the several TeV region. {In Tables \ref{tab1} and \ref{tab2}}  we exhibit the results for two benchmark points where we assume
universality and take the scalar masses and the trilinear couplings to be all equal. Table \ref{tab1} exhibits the result of
the computation for $\Delta a_\mu$ where
individual contributions arising from the chargino exchange, neutralino exchange, $W$ exchange and $Z$ exchange are listed.
 The entries exhibit the contributions over and above what one expects from the standard model and so
 the
 entries for the
 $W$ and $Z$ exchanges show a null value. Thus the entire contribution in this case arises from the chargino and the neutralino
 exchange and their sum gives a value ${\cal O}(10^{-11})$ which is two orders of magnitude smaller than the
 experimental result of Eq. (\ref{eq1}). A very similar analysis is given in Table \ref{tab2} for $\Delta a_e$ where again the contribution
 to $\Delta a_e$ arises from the exchange of charginos and neutralinos and their sum is ${\cal O}(10^{-16})$ which is three
 orders of magnitude smaller than the result of Eq. (\ref{eq2}). Thus with a high scale of the scalar masses one cannot explain
 the results of Eq. (\ref{eq1}) and Eq. (\ref{eq2}).

We turn now to the analysis within the extended MSSM with a vector like leptonic generation.
As in the analysis within MSSM here also we assume the universality of the soft parameters so that
 we set $m_{0}^{2}=\tilde{M}_{\tau L}^{2}=\tilde{M}_{E}^{2}=\tilde{M}_{\tau}^{2}=\tilde{M}_{\chi}^{2}=\tilde{M}_{\mu L}^{2}=\tilde{M}_{\mu}^{2}=\tilde{M}_{e L}^{2}=\tilde{M}_{e}^{2}=\tilde{M}_{4L}^{2}=\tilde{M}_{4}^{2}$ and $A_{0}=A_{\tau}=A_{E}=A_{\mu}=A_{e}=A_{4\ell}$
 in the computation of the charged slepton mass squared matrix. Similarly we assume $m_{0}^{\tilde{\nu}^{2}}=\tilde{M}_{N}^{2}=\tilde{M}_{\nu_{\tau}}^{2}=\tilde{M}_{\nu_{\mu}}^{2}=\tilde{M}_{\nu_{e}}^{2}=\tilde{M}_{4L}^{2}=\tilde{M}_{\nu 4}^{2}$ and $A_{\nu_{\tau}}= A_{\nu_{\mu}}= A_{\nu_{e}}=A_{N}=A_{4\nu}=A_{0}^{\tilde{\nu}}$ for the computation of the sneutrino mass squared matrix (see Appendix).
 The contributions from the chargino exchange, the neutralino exchange, and the $W$ and $Z$ exchange are listed in Table \ref{tab1}
 and Table \ref{tab2} for two benchmark points. In this case the $W$ boson and the $Z$ boson exchange contributions
 are non-vanishing and the contributions listed are those over and above what one expects in the standard model.
 As in the MSSM case here also one finds that the contributions from the chargino exchange and from the neutralino exchange  fall significantly below the experimental results of Eq. (\ref{eq1}) and Eq. (\ref{eq2}). However, in this case including the
  contributions from the $W$ exchange and from the $Z$ boson exchange one finds that consistency with Eq. (\ref{eq1}) and Eq. (\ref{eq2})
  is  achieved. {At the same time one has the Higgs boson mass in the model for both
  benchmarks (a) and (b) at  $\sim 125$ GeV   consistent with the experimental measurements
by ATLAS~\cite{Aad:2012tfa} and by CMS ~\cite{Chatrchyan:2012xdj}. Here the loop correction 
{that gives mass to the Higgs boson} 
comes
from the MSSM sector while the  extra vector like leptonic generation makes a negligible contribution. }

  In the analysis {of $\Delta a_\mu$ and $\Delta a_e$}   
   the exchange of both the sequential leptons and the mirrors play a role {with the
  mirror exchange being the more dominant.}
   The analysis requires diagonalization of a $5\times 5$ mass matrix in the charged lepton-charged mirror lepton sector
  and diagonalization of a $5\times 5$ mass matrix in the neutrino-mirror  neutrino sector. Parameter choices are made to
  ensure that the eigenvalues in the charged lepton sector give the desired experimental values for $e$, $\mu$ and
  $\tau$ along with two additional masses, one for the sequential fourth generation lepton and the other for the mirror charged
  lepton.  Their values are listed in Table \ref{tab3} for the case of two benchmark points (a) and (b).
  A similar analysis holds for the neutrino-mirror neutrino sector where we get two additional eigenvalues, one for the
  fourth generation neutrino and the other for the mirror neutrino. Their values are also listed in Table \ref{tab3} for
  two benchmark points. {The analysis also requires diagonalization of a $10\times 10$ matrix in the charged  slepton
  and charged  mirror slepton sector, as well as diagonalization of a $10\times 10$ matrix in the sneutrino
  and the  mirror sneutrino sector.}
    \\

\begin{table}[H]
\begin{center}
\begin{tabular}{l c c c c c}
  \hline \hline
   \multicolumn{2}{c}{}   & \multicolumn{2}{c}{(a)} & \multicolumn{2}{c}{(b)} \\
      \hline
  \multicolumn{2}{c}{Contribution} & MSSM & Vectorlike & MSSM & Vectorlike \\
  \hline
  {Chargino} & $a_{\mu}^{\chi^{\pm}}$ & $+1.68\times 10^{-11}$ & $+1.07\times 10^{-11}$ & $+1.68\times 10^{-11}$ & $-8.54\times 10^{-11}$ \\
  {Neutralino} & $a_{\mu}^{\chi^{0}}$ & $-3.09\times 10^{-13}$ & $-1.50\times 10^{-12}$ & $-3.09\times 10^{-13}$ & $-6.58\times 10^{-13}$ \\
  {W Boson} & $a_{\mu}^{W}$ & $0$ & $+1.53\times 10^{-9}$ & $0$ & $+2.56\times 10^{-9}$ \\
  {Z Boson} & $a_{\mu}^{Z}$ & $0$ & $+5.12\times 10^{-10}$ & $0$ & $+8.76\times 10^{-10}$ \\
  {Total} & $\Delta a_{\mu}$ & $+1.65\times 10^{-11}$ & $+2.05\times 10^{-9}$ & $+1.65\times 10^{-11}$ & $+3.35\times 10^{-9}$ \\
  \hline
\end{tabular}
\caption{The contribution of the vectorlike multiplet vs the contribution from the MSSM
sector to the anomalous magnetic moments of the muon  for two illustrative benchmark
points (a) and (b). They are: (a) $m_N=5$, $m_{4\ell}=450$, $|f'_3|=0.62$, $|f''_{3}|=6.62\times 10^{-3}$, $|f'_{4}|=20$, $|h_6|=230$, $|h_8|=730$ and (b) $m_N=200$, $m_{4\ell}=250$, $|f'_3|=0.73$, $|f''_{3}|=5.23\times 10^{-3}$, $|f'_{4}|=30$, $|h_6|=66$, $|h_8|=180$ . Other parameters have the values $\tan\beta=15$, $m_{0}=m_{0}^{\tilde{\nu}}=5000$, $|A_{0}^{\tilde{\nu}}|=|A_{0}|=6000$, $|m_{1}|=224$, $|m_{2}|=407$, $|\mu|=2124$, $m_{E}=320$, $m_{\nu 4}=350$, $m_{h^{0}}=124.66$, $|f_{3}|=1\times 10^{-4}$, $|f_{4}|=1\times 10^{-5}$, $|f''_{4}|=38$, $|f_{5}|=1\times 10^{-4}$, $|f'_{5}|=5.0 \times 10^{-4}$, $|f''_{5}|=3.0\times 10^{-3}$, $|h_7|=34$, $\alpha_{A_{0}}=\pi$, $\alpha_{A_{0}^{\tilde{\nu}}}=\pi$, $\xi_{1}=\xi_{2}=\theta_{\mu}=\chi_3=\chi'_3=\chi''_3=\chi_4=\chi'_4=\chi''_4=\chi_5=\chi'_5=\chi''_5=\chi_6=\chi_7=\chi_8=0$.
For the MSSM analysis the following parameters were used for both cases (a) and (b):
The scalar masses are taken to be universal with $m_0= 5000$ and the trilinear coupling is taken to be
universal $A_0=-600$. Other inputs are:
$\chi_1^{\pm}= 223$, $\chi_1^0= 220$, $\chi_2^{\pm}= \chi_2^0= 440$,  $-\chi^0_3=\chi_4^0= 214$,
$\mu=214$.
All masses are in GeV and phases in rad.}
 \label{tab1}
\end{center}
\end{table}

\begin{table}[H]
\begin{center}
\begin{tabular}{l c c c c c}
  \hline \hline
   \multicolumn{2}{c}{}   & \multicolumn{2}{c}{(a)} & \multicolumn{2}{c}{(b)} \\
      \hline
  \multicolumn{2}{c}{Contribution} & MSSM & Vectorlike & MSSM & Vectorlike \\
  \hline
  {Chargino} & $a_{e}^{\chi^{\pm}}$ & $+3.92\times 10^{-16}$ & $-2.88\times 10^{-16}$ & $+3.92\times 10^{-16}$ & $-6.31\times 10^{-15}$ \\
      {Neutralino}     & $a_{e}^{\chi^{0}}$ & $-7.25\times 10^{-18}$ & $-1.69\times 10^{-16}$ & $-7.25\times 10^{-18}$ & $-3.12\times 10^{-17}$ \\
        {W Boson}                    & $a_{e}^{W}$ & $0$ & $+1.99\times 10^{-13}$ & $0$ & $+1.71\times 10^{-13}$ \\
       {Z Boson}                    & $a_{e}^{Z}$ & $0$ & $+5.89\times 10^{-14}$ & $0$ & $+5.11\times 10^{-14}$ \\
           {Total}               & $\Delta a_{e}$ & $+3.85\times 10^{-16}$ & $+2.58\times 10^{-13}$ & $+3.85\times 10^{-16}$ & $+2.16\times 10^{-13}$ \\
  \hline
\end{tabular}
\caption{The contribution of the vectorlike multiplet vs the contribution from the MSSM
sector to the anomalous magnetic moments of the electron for two illustrative benchmark
points (a) and (b) as given in table \ref{tab1}.}
\label{tab2}
\end{center}
\end{table}

\begin{table}[H] \centering
\begin{tabular}{lrc}
\hline \hline
& \multicolumn{2}{c}{Mass Spectrum (GeV)} \\
\hline
Particles & (a) & (b) \\ \hline
Mirror Neutrino & 208 & 207  \\
Fourth Sequential Neutrino & 816 & 395  \\
Mirror Lepton & 253 & 349  \\
Fourth Sequential Lepton & 545 & 226  \\ \hline
\end{tabular}
\caption{The mass of the heavy particles obtained after diagonalizing the lepton and neutrino mass matrices for benchmark points (a) and (b) of Table~\ref{tab1}.} \label{tab3}
\end{table}

\begin{figure}[H]
\begin{center}
{\rotatebox{0}{\resizebox*{9cm}{!}{\includegraphics{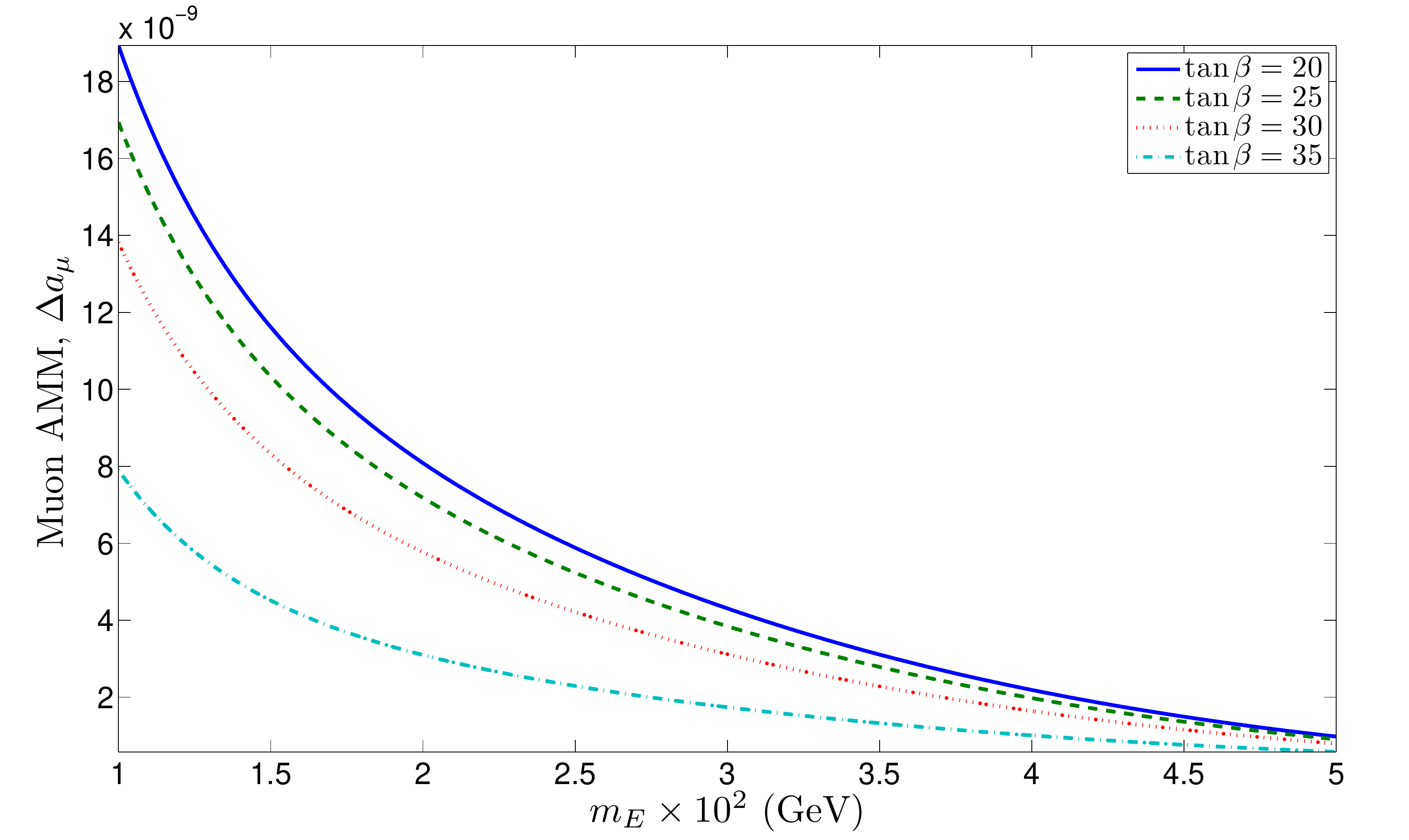}}\hglue5mm}}
\caption{$\Delta a_{\mu}$ as a function of $m_E$  when $\tan\beta$ = 20, 25, 30, 35. Other parameters are $m_{0}=m_{0}^{\tilde{\nu}}=5000$, $|A_{0}^{\tilde{\nu}}|=|A_{0}|=6000$, $|m_{1}|=224$, $|m_{2}|=407$, $|\mu|=2124$, $m_{4\ell}=250$, $m_N=300$, $m_{\nu 4}=350$, $m_{h^{0}}=124.66$, $|f_{3}|=1\times 10^{-5}$, $|f'_3|=8.18$, $|f''_{3}|=4.32\times 10^{-2}$, $|f_{4}|=1\times 10^{-3}$, $|f'_{4}|=3.61$, $|f''_{4}|=3.85$, $|f_{5}|=1\times 10^{-4}$, $|f'_{5}|=5.0 \times 10^{-4}$, $|f''_{5}|=3.0\times 10^{-6}$, $|h_6|=10$, $|h_7|=19$, $|h_8|=10$, $\alpha_{A_{0}}=\pi$, $\alpha_{A_{0}^{\tilde{\nu}}}=\pi$, $\xi_{1}=\xi_{2}=\theta_{\mu}=\chi_3=\chi'_3=\chi''_3=\chi_4=\chi'_4=\chi''_4=\chi_5=\chi'_5=\chi''_5=\chi_6=\chi_7=\chi_8=0$. All masses are in GeV and phases in rad.}
\label{fig3}
\end{center}
\end{figure}

We discuss now some further features of the analysis which includes the vector like leptonic generation.
In Figure~\ref{fig3} we show the variation of  $\Delta a_{\mu}$ as a function of $m_E$ the mass of the mirror lepton
 as given by Eq.~(\ref{me}), for four $\tan\beta$ values. A remarkable feature of this graph is the dependence on $\tan\beta$
 it exhibits. Notice that for a fixed $m_E$, $\Delta a_\mu$ decreases for increasing values of $\tan\beta$ as $\tan\beta$ varies
 from $20$ to $35$. Now we recall that the Yukawa coupling of a charged lepton has a $1/\cos\beta$ dependence and
 as a consequence the contribution of the charged lepton to $\Delta a_\mu$ becomes larger for larger 
 $\tan\beta$ {which is a  well known result.} 
 However, the Yukawa coupling of the mirror lepton goes like $1/\sin\beta$~\cite{Ibrahim:2008gg}
 and so $\Delta a_\mu$   decreases for larger
 values of $\tan\beta$. 
 This feature explains the $\tan\beta$ dependence in Figure ~\ref{fig3}. It also shows that the
 $W$ and $Z$ exchange contributions in this case are being controlled by exchange of the mirror particles. A very similar
 dependence on $\tan\beta$ is exhibited by $\Delta a_e$. \\

\begin{figure}[H]
\begin{center}
{\rotatebox{0}{\resizebox*{8cm}{!}{\includegraphics{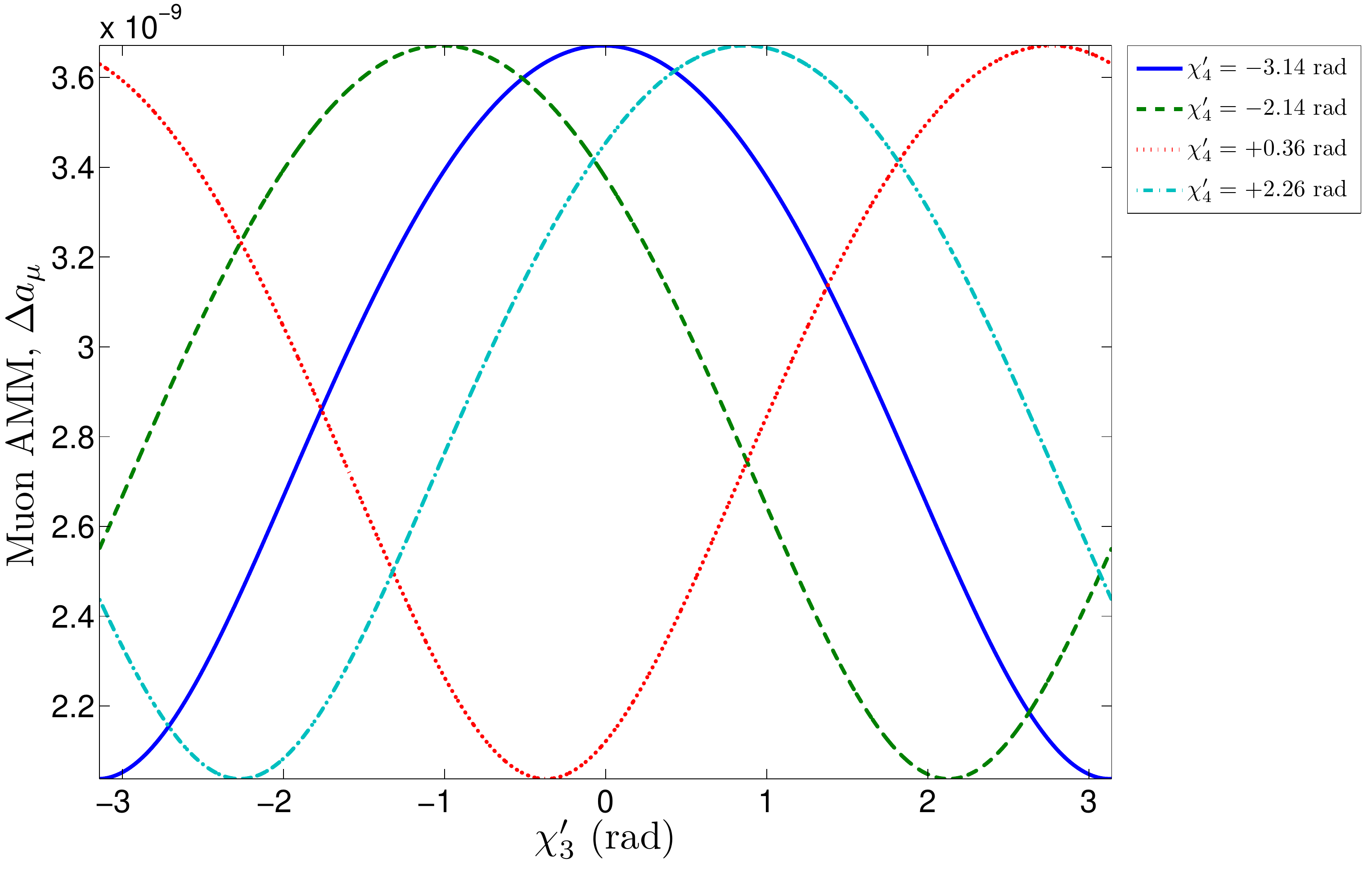}}\hglue5mm}}
{\rotatebox{0}{\resizebox*{8cm}{!}{\includegraphics{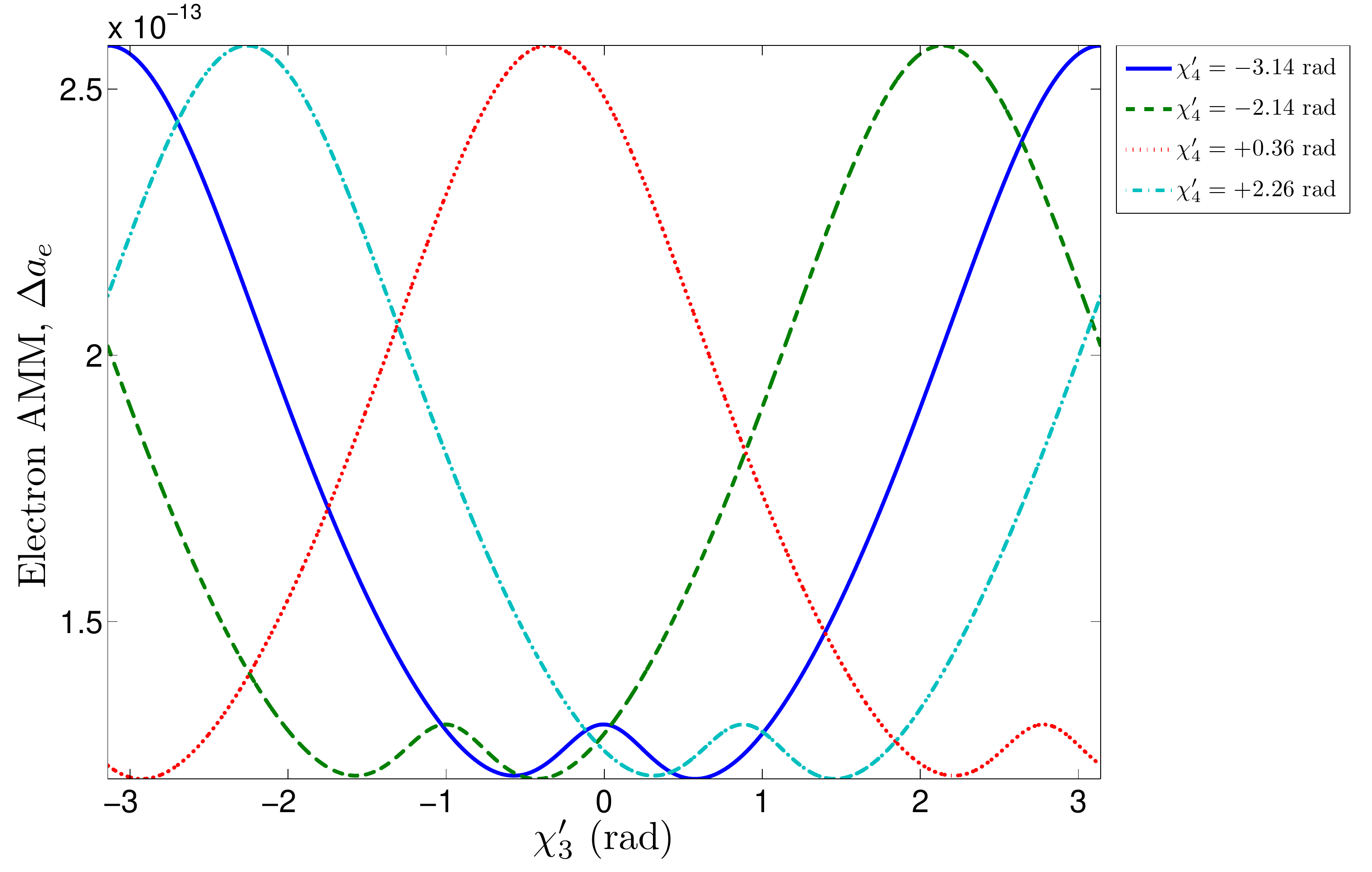}}\hglue5mm}}
\caption{ $\Delta a_{\mu}$ (left panel) and  $\Delta a_{e}$ (right panel) as a function of $\chi'_3$ in the range $[-\pi,+\pi]$ when $\chi'_4=-3.14,-2.14,+0.36,+2.26$. Other parameters are $\tan\beta=15$, $m_0=m_{0}^{\tilde{\nu}}=5000$, $|A_{0}^{\tilde{\nu}}|=|A_{0}|=6000$, $|m_{1}|=224$, $|m_{2}|=407$, $|\mu|=2124$, $m_{N}=5$, $m_{\nu 4}=350$, $m_{E}=320$, $m_{4\ell}=450$, $m_{h^{0}}=124.66$, $|f_{3}|=1\times 10^{-4}$, $|f'_{3}|=0.62$, $|f''_{3}|=6.62\times 10^{-3}$, $|f_{4}|=1\times 10^{-5}$, $|f'_{4}|=20$, $|f''_{4}|=38$, $|f_{5}|=1\times 10^{-4}$, $|f'_{5}|=5.0 \times 10^{-4}$, $|f''_{5}|=3.0\times 10^{-6}$, $|h_6|=230$, $|h_7|=34$, $|h_8|=730$, $\alpha_{A_{0}}=\pi$, $\alpha_{A_{0}^{\tilde{\nu}}}=\pi$, $\xi_{1}=\xi_{2}=\theta_{\mu}=\chi_3=\chi''_3=\chi_4=\chi''_4=\chi_5=\chi'_5=\chi''_5=\chi_6=\chi_7=\chi_8=0$. All masses are in GeV and phases in rad.}
\label{fig4}
\end{center}
\end{figure}

The anomalous magnetic moments are quite sensitive to CP phases as first demonstrated in the analysis of ~\cite{Ibrahim:1999hh,Ibrahim:1999aj,Ibrahim:2001ym} for the case of CP phases that arise in $N=1$ supergravity ~\cite{Ibrahim:1999hh,Ibrahim:2001ym}
and more generally for the case of MSSM~\cite{Ibrahim:1999aj}. In those analyses it was also found that large CP phases
could be made consistent with the {experimental  constraints on the EDMs}   by 
the cancellation mechanism~\cite{Ibrahim:1997gj,Falk:1998pu,Ibrahim:1998je,Brhlik:1998zn,Ibrahim:1999af}.
In the present
analysis the contribution from the MSSM sector is suppressed and the dominant contribution arises from the $W$ and $Z$
exchanges. For the case of three generations this sector does not have any CP phases in the leptonic sector. However,
the extended MSSM with a vector like leptonic multiplet allows for CP phases which cannot be removed by field redefinitions.
It is of interest then to discuss the dependence of $\Delta a_\mu$ and $\Delta a_e$ on the CP phases that arise in
the extended MSSM. We discuss now the dependence of $\Delta a_\mu$ and $\Delta a_e$ on such phases.
In Fig. (\ref{fig4}) we exhibit the dependence of $\Delta a_\mu$ and $\Delta a_e$ on $\chi'_3$,
which is the phase of $f_3'$ (see Appendix). A sharp dependence on $\chi_3'$ is seen for both
$\Delta a_\mu$ and for $\Delta a_e$.  A very similar sensitivity to the CP phase $\chi_6$ which is the phase of
$h_6$ (see Appendix) is exhibited in Fig. (\ref{fig5}).  To explore further the sensitivity of
$\Delta a_\mu$ and of $\Delta a_e$ to parameters in the vector like sector we exhibit in
Fig. (\ref{fig6}) the dependence of $\Delta a_\mu$ and $\Delta a_e$ on $h_6$ which is the
co-efficient of the term  $\epsilon_{ij}  \hat\chi^c{^{i}}\hat\psi_{4 L}^{j}$ in the superpotential
(see Eq. (\ref{5})). 
One can see in Fig. (\ref{fig6}) the strong dependence of $\Delta a_\mu$ and $\Delta a_e$  on $h_6$.
In the analyses given so far both $\Delta a_\mu$ and $\Delta a_e$
have very significant dependence on the parameters arising from inclusion of the vector like sector.
However, there are parameters which  affect $\Delta a_e$ and $\Delta a_\mu$  differently.
{This is the case for $|f^{''}_3|$. { Here as seen in the left panel of Fig. (\ref{fig7})
$\Delta a_e$ is a sensitive function of $|f^{''}_3|$ but not so for the case for $\Delta a_\mu$ (not exhibited) because of
its much larger size.
Finally we note that even if the SUSY scale lies in the PeV region, the contributions from the $W$ and $Z$ exchange
arising from Fig. (\ref{fig2}) survive while the diagrams of Fig. (\ref{fig1}) give a vanishingly small contribution.
This is illustrated in the right panel of Fig. (\ref{fig7}).
}}

\begin{figure}[H]
\begin{center}
{\rotatebox{0}{\resizebox*{8cm}{!}{\includegraphics{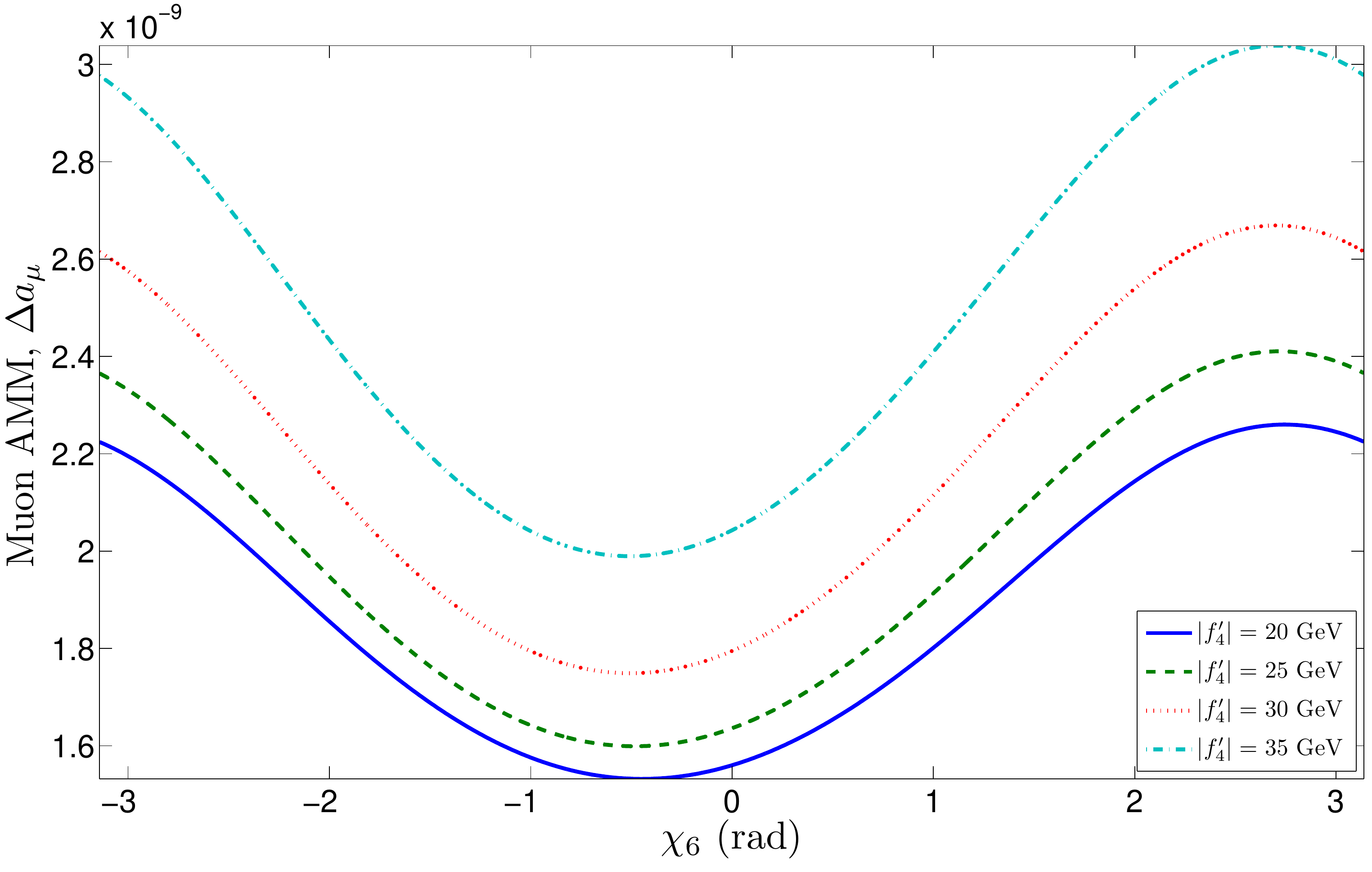}}\hglue5mm}}
{\rotatebox{0}{\resizebox*{8cm}{!}{\includegraphics{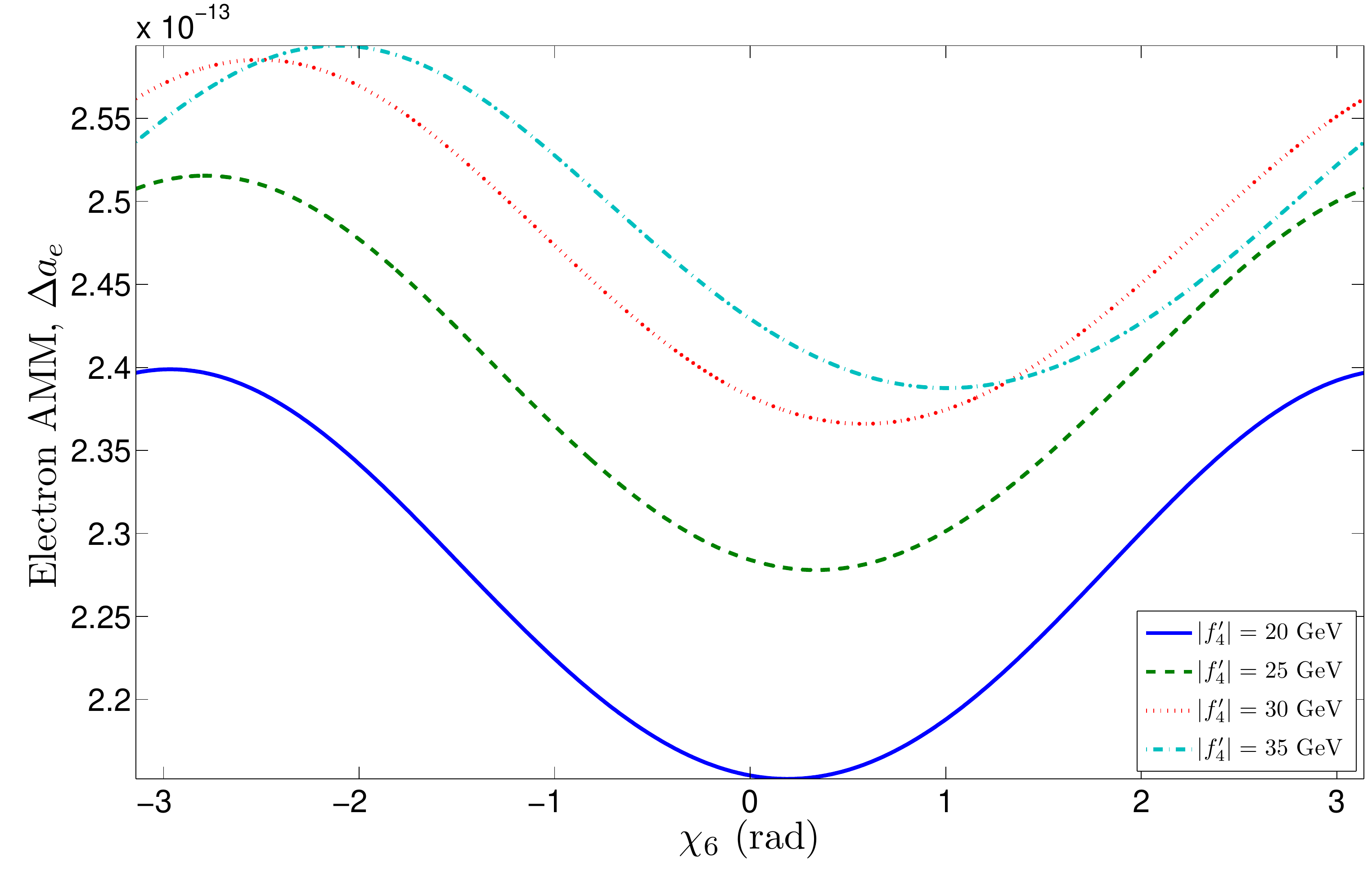}}\hglue5mm}}
\caption{$\Delta a_{\mu}$ (left panel) and $\Delta a_{\mu}$ (right panel) as a function of $\chi_6$ in the range $[-\pi,+\pi]$ when $|f'_4|=20,25,30,35$. Other parameters are $\tan\beta=15$, $m_0=m_{0}^{\tilde{\nu}}=5000$, $|A_{0}^{\tilde{\nu}}|=|A_{0}|=6000$, $|m_{1}|=224$, $|m_{2}|=407$, $|\mu|=2124$, $m_{N}=5$, $m_{\nu 4}=350$, $m_{E}=320$, $m_{4\ell}=450$, $m_{h^{0}}=124.66$, $|f_{3}|=1\times 10^{-4}$, $|f'_{3}|=0.627$, $|f''_{3}|=6.605\times 10^{-3}$, $|f_{4}|=1\times 10^{-3}$, $|f''_{4}|=38$, $|f_{5}|=1\times 10^{-4}$, $|f'_{5}|=5.0 \times 10^{-4}$, $|f''_{5}|=3.0\times 10^{-6}$, $|h_6|=230$, $|h_7|=34$, $|h_8|=730$, $\alpha_{A_{0}}=\pi$, $\alpha_{A_{0}^{\tilde{\nu}}}=\pi$, $\xi_{1}=\xi_{2}=\theta_{\mu}=\chi_3=\chi_4=\chi_5=\chi'_5=\chi''_5=0$, $\chi'_3=2.96$, $\chi''_3=-1.54$, $\chi'_4=2.86$, $\chi''_4=1.46$, $\chi_7=-2.94$, $\chi_8=0.6$. All masses are in GeV and phases in rad.}
\label{fig5}
\end{center}
\end{figure}

\begin{figure}[H]
\begin{center}
{\rotatebox{0}{\resizebox*{8cm}{!}{\includegraphics{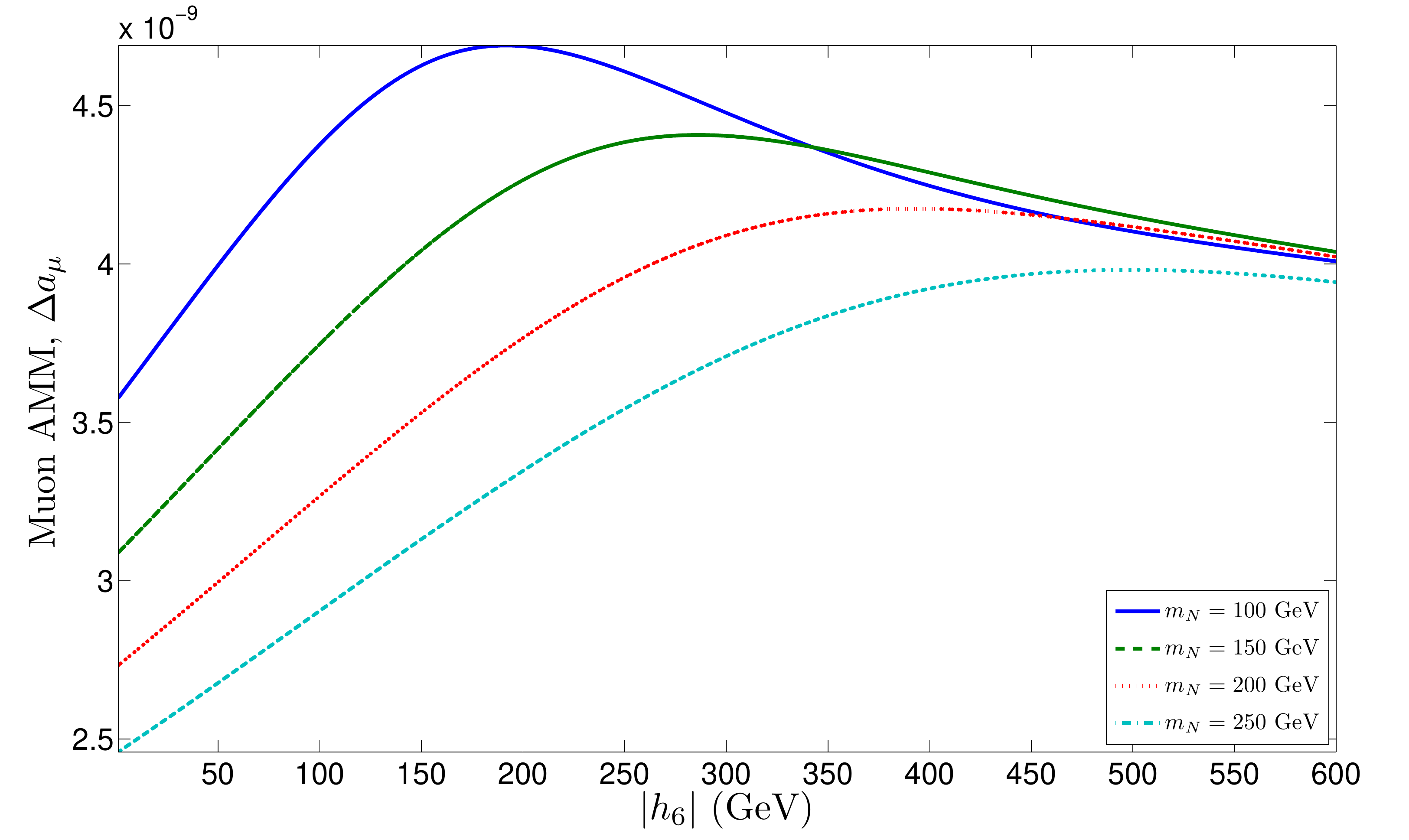}}\hglue5mm}}
{\rotatebox{0}{\resizebox*{8cm}{!}{\includegraphics{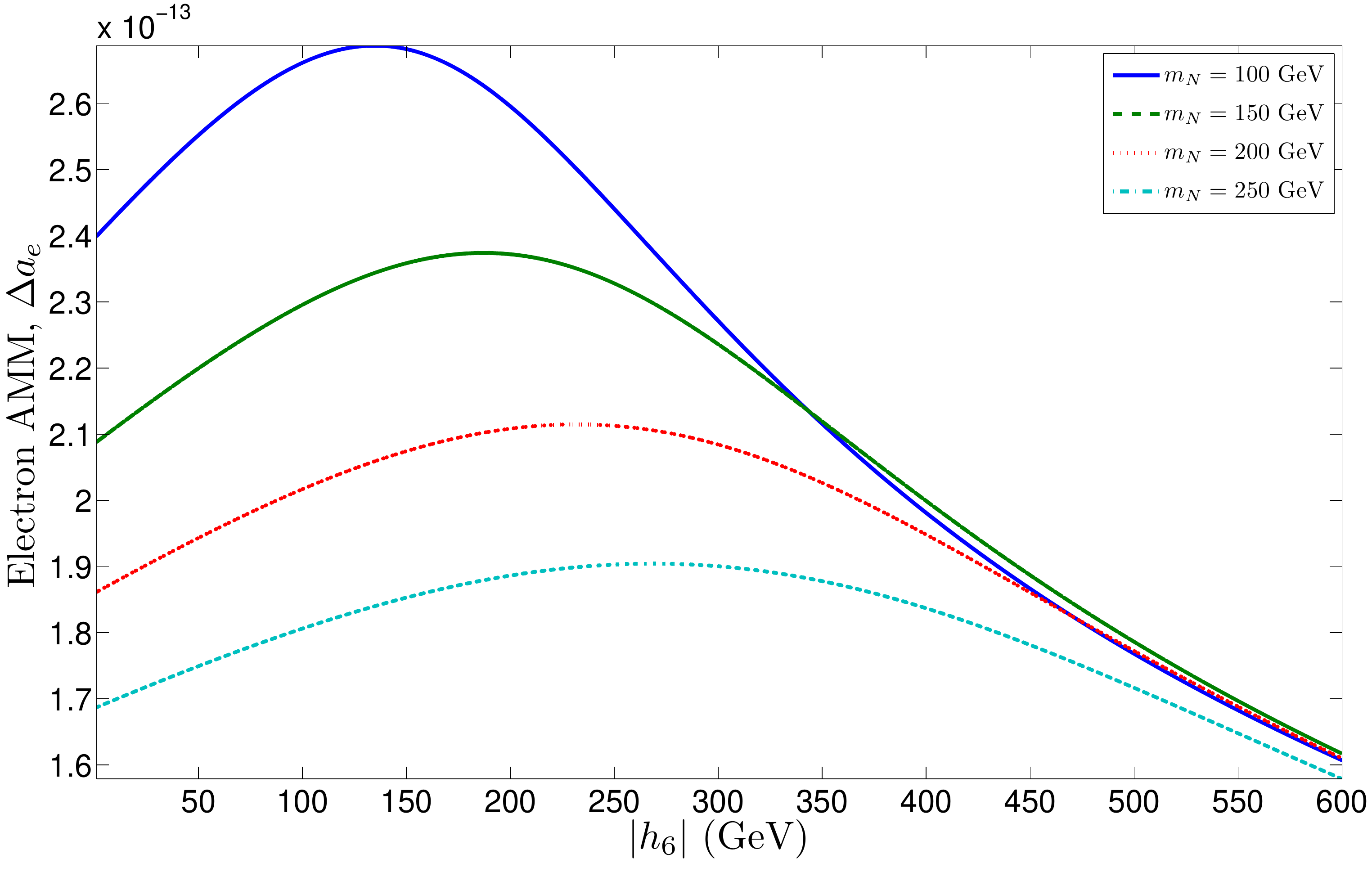}}\hglue5mm}}
\caption{ $\Delta a_{\mu}$ (left panel) and $\Delta a_{e}$ (right panel) as a function of $|h_6|$ when $m_N=100,150,200,250$. Other parameters are $\tan\beta=15$, $m_0=m_{0}^{\tilde{\nu}}=5000$, $|A_{0}^{\tilde{\nu}}|=|A_{0}|=6000$, $|m_{1}|=224$, $|m_{2}|=407$, $|\mu|=2124$, $m_{\nu 4}=350$, $m_{E}=320$, $m_{4\ell}=250$, $m_{h^{0}}=124.66$, $|f_{3}|=1\times 10^{-5}$, $|f'_{3}|=0.73$, $|f''_{3}|=5.23\times 10^{-3}$, $|f_{4}|=1\times 10^{-3}$, $|f'_{4}|=20$, $|f''_{4}|=38$, $|f_{5}|=1\times 10^{-4}$, $|f'_{5}|=5.0 \times 10^{-4}$, $|f''_{5}|=3.0\times 10^{-6}$, $|h_7|=34$, $|h_8|=180$, $\alpha_{A_{0}}=\pi$, $\alpha_{A_{0}^{\tilde{\nu}}}=\pi$, $\xi_{1}=\xi_{2}=\theta_{\mu}=\chi_3=\chi_4=\chi_5=\chi'_5=\chi''_5=0$, $\chi'_3=2.96$, $\chi''_3=-1.54$, $\chi'_4=2.86$, $\chi''_4=1.46$, $\chi_6=3.06$, $\chi_7=-2.94$, $\chi_8=0.6$. All masses are in GeV and phases in rad.}
\label{fig6}
\end{center}
\end{figure}

\begin{figure}[H]
\begin{center}
{\rotatebox{0}{\resizebox*{8cm}{!}{\includegraphics{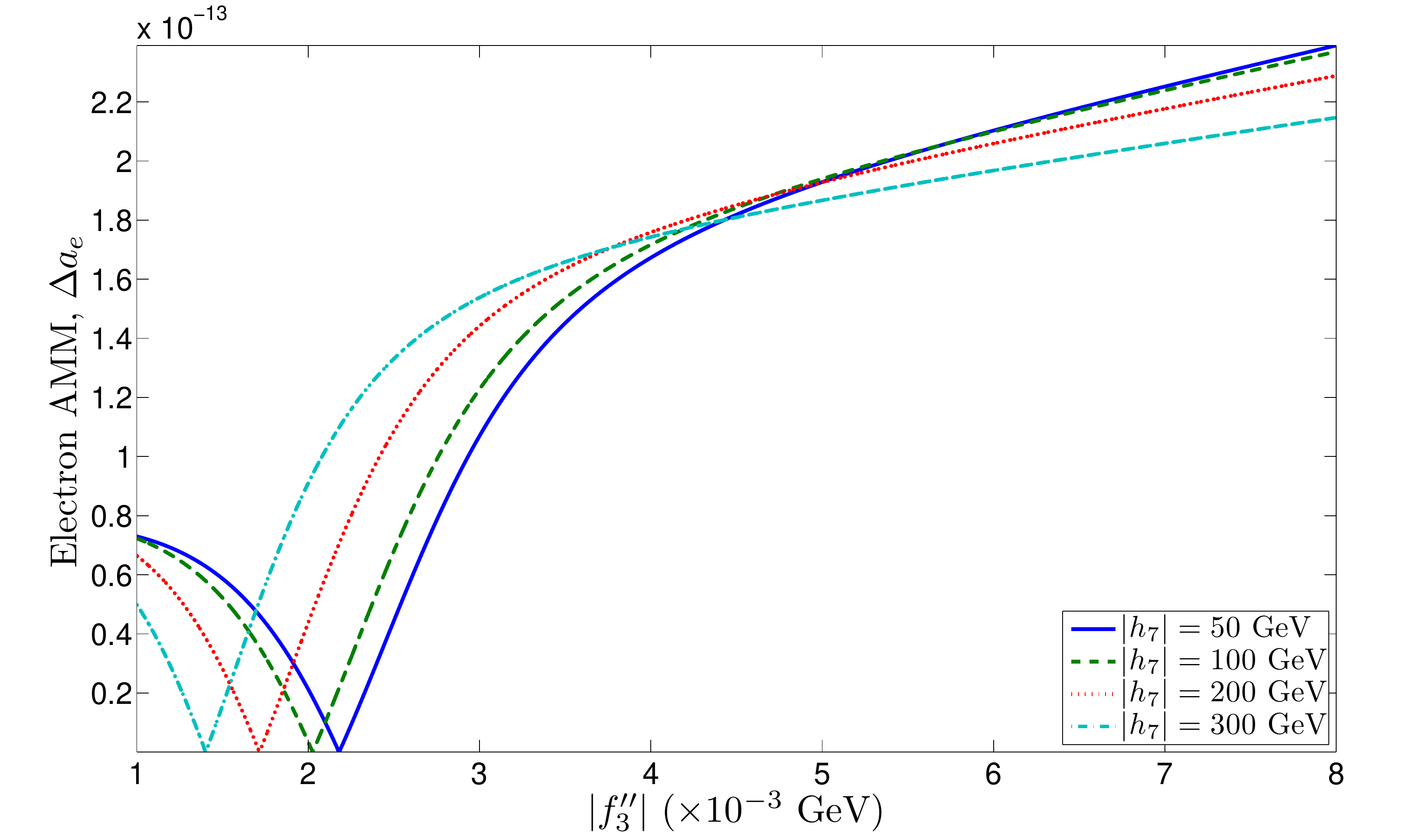}}\hglue5mm}}
{\rotatebox{0}{\resizebox*{8cm}{!}{\includegraphics{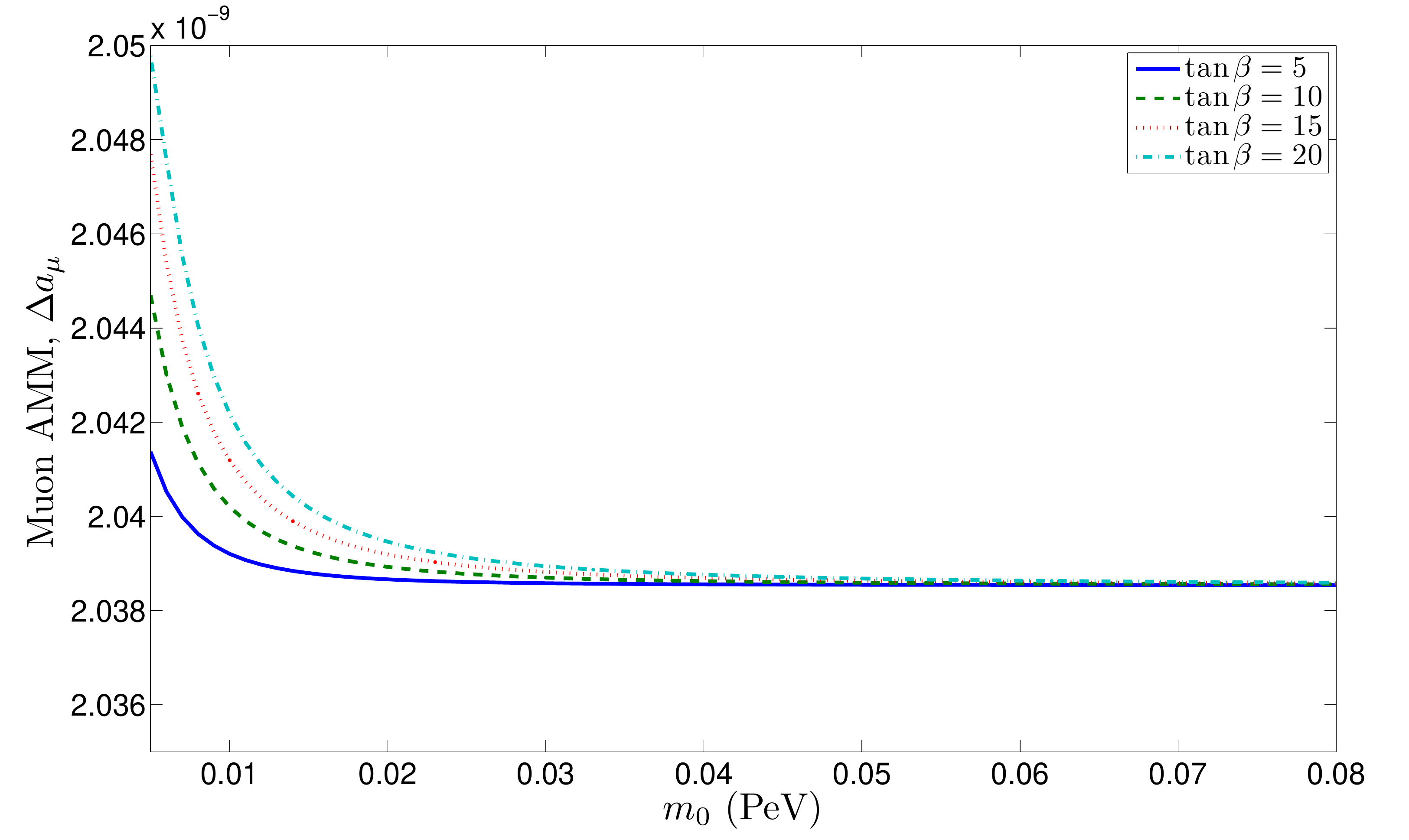}}\hglue5mm}}
\caption{{Left panel: Variation of $\Delta a_{e}$ as a function of $|f''_3|$ for four values of $|h_7|$ when $|h_7|=50,100,200,300$. Other parameters are $\tan\beta=15$, $m_0=m_{0}^{\tilde{\nu}}=5000$, $|A_{0}^{\tilde{\nu}}|=|A_{0}|=6000$, $|m_{1}|=224$, $|m_{2}|=407$, $|\mu|=2124$, $m_N=200$, $m_{\nu 4}=350$, $m_{E}=320$, $m_{4\ell}=250$, $m_{h^{0}}=124.66$, $|f_{3}|=1\times 10^{-5}$, $|f'_{3}|=0.73$, $|f_{4}|=1\times 10^{-3}$, $|f'_{4}|=20$, $|f''_{4}|=38$, $|f_{5}|=1\times 10^{-4}$, $|f'_{5}|=5.0 \times 10^{-4}$, $|f''_{5}|=3.0\times 10^{-6}$, $|h_6|=66$, $|h_8|=180$, $\alpha_{A_{0}}=\pi$, $\alpha_{A_{0}^{\tilde{\nu}}}=\pi$, $\xi_{1}=\xi_{2}=\theta_{\mu}=\chi_3=\chi_4=\chi_5=\chi'_5=\chi''_5=0$, $\chi'_3=2.96$, $\chi''_3=-1.54$, $\chi'_4=2.86$, $\chi''_4=1.46$, $\chi_6=3.06$, $\chi_7=-2.94$, $\chi_8=0.6$. All masses are in GeV and phases in rad.}  {Right panel: A plot of $\Delta a_\mu$ as a function of the common scalar mass $m_0$ exhibiting a residual
correction $\Delta a_\mu$ even when $m_0$ lies in the PeV region.   The parameters used in the plot are for benchmark (a) in Table (\ref{tab1}).}
 }
\label{fig7}
\end{center}
\end{figure}

\section{4. Conclusion\label{secconclu}}
The  Higgs boson mass measurement at 126 GeV indicates a high SUSY scale, and specifically a high
scale for the scalar masses. If the scalar masses are all heavy, the contribution to the leptonic moments
and specifically to $\Delta a_{\ell}= a_{\ell}^{\rm exp}- a_{\ell}^{\rm SM}$ becomes negligible in this case.
In this work we have  investigated leptonic $g-2$ moments within an extended MSSM model with an extra
vector like generation and CP phase dependent couplings. It is found that one can
achieve consistency with the experimental measurements of $\Delta a_\mu$ and $\Delta {a_e}$
under the constraint of the Higgs boson mass. The dependence of the moments on CP phases
from the new sector are also investigated and shown to have a very sensitive dependence.
Further, it is shown that $\Delta a_{\mu}$ and $\Delta a_e$ will be non-vanishing even when
the SUSY scale lies in the PeV region. {The model presented here can be made UV complete
by including a full generation of vector like matter including both quarks and leptons.
Finally we note that the work presented here has some
overlap with \cite{Nishida:2016lyk} which appeared after  this work was finished.}\\

\noindent
{\em Acknowledgments}:
This research was supported in part by the NSF Grant PHY-1314774.\\

\section{Appendix on the extended MSSM with a vector like leptonic generation}

In this Appendix we define the notation for the vector generation and their
properties under $SU(3)_C\times SU(2)_L \times U(1)_Y$.
For the four sequential families we use the notation
\beqn
\psi_{iL}\equiv \left(
\begin{array}{c}
 \nu_{iL}\\
 \ell_{iL}
\end{array}\right) \sim(1,2,- \frac{1}{2}), \ell^c_{iL}\sim (1,1,1), \nu^c_{iL}\sim (1,1,0),
\eeqn
where the last entry on the right hand side of each $\sim$ is the value of the hypercharge
 $Y$ defined so that $Q=T_3+ Y$ and we have included in our analysis the singlet field
 $\nu^c_i$, where $i$ runs from $1-4$.
For the mirrors we use the notation
\beqn
\chi^c\equiv \left(
\begin{array}{c}
 E_{\mu L}^c\\
 N_L^c
\end{array}\right)
\sim(1,2,\frac{1}{2}), E_{\mu L}\sim (1,1,-1), N_L\sim (1,1,0).
\eeqn
The main difference between the leptons and the mirrors is that while the
 leptons have $V-A$ type interactions with $SU(2)_L\times U(1)_Y$
  gauge bosons the mirrors have  $V+A$ type interactions.

We assume that the mirrors of the vector like generation escape acquiring mass at the GUT scale and remain
light down to the electroweak scale
where the superpotential of the model for the lepton part  may be written  in the form

\begin{align}
W&= -\mu \epsilon_{ij} \hat H_1^i \hat H_2^j+\epsilon_{ij}  [f_{1}  \hat H_1^{i} \hat \psi_L ^{j}\hat \tau^c_L
 +f_{1}'  \hat H_2^{j} \hat \psi_L ^{i} \hat \nu^c_{\tau L}
+f_{2}  \hat H_1^{i} \hat \chi^c{^{j}}\hat N_{L}
 +f_{2}'  \hat H_2^{j} \hat \chi^c{^{i}} \hat E_{ L} \nonumber \\
&+ h_{1}   \hat H_1^{i} \hat\psi_{\mu L} ^{j}\hat\mu^c_L
 +h_{1}'  \hat  H_2^{j} \hat\psi_{\mu L} ^{i} \hat\nu^c_{\mu L}
+ h_{2}  \hat  H_1^{i} \hat\psi_{e L} ^{j}\hat e^c_L
 +h_{2}'  \hat  H_2^{j} \hat\psi_{e L} ^{i} \hat\nu^c_{e L}
+y_{5}  \hat  H_1^{i} \hat\psi_{4 L} ^{j} \hat\ell^c_{4 L}
+y_{5} '  \hat  H_2^{j} \hat\psi_{4 L} ^{i} \hat\nu^c_{4 L}
] \nonumber \\
&+ f_{3} \epsilon_{ij}  \hat\chi^c{^{i}}\hat\psi_L^{j}
 + f_{3}' \epsilon_{ij}  \hat\chi^c{^{i}}\hat\psi_{\mu L}^{j}
 + f_{4} \hat\tau^c_L \hat E_{ L}  +  f_{5} \hat\nu^c_{\tau L} \hat N_{L}
 + f_{4}' \hat\mu^c_L \hat E_{ L}  +  f_{5}' \hat\nu^c_{\mu L} \hat N_{L} \nonumber \\
&+ f_{3}'' \epsilon_{ij}  \hat\chi^c{^{i}}\hat\psi_{e L}^{j}x
 + f_{4}'' \hat e^c_L \hat E_{ L}  +  f_{5}'' \hat\nu^c_{e L} \hat N_{L}\
+ h_6 \epsilon_{ij}  \hat\chi^c{^{i}}\hat\psi_{4 L}^{j}
+h_7  \hat \ell^c_{4L} \hat E_{ L}
+h_8  \hat \nu^c_{4L} \hat N_{ L}
 ,
 \label{5}
\end{align}
where  $\hat ~$ implies superfields,  $\hat\psi_L$ stands for $\hat\psi_{3L}$, $\hat\psi_{\mu L}$ stands for $\hat\psi_{2L}$
and  $\hat\psi_{e L}$ stands for $\hat\psi_{1L}$.

The mass terms for the neutrinos, mirror neutrinos,  leptons and  mirror leptons arise from the term
\beq
{\cal{L}}=-\frac{1}{2}\frac{\partial ^2 W}{\partial{A_i}\partial{A_j}}\psi_ i \psi_ j+\text{H.c.}
\label{6}
\eeq
where $\psi$ and $A$ stand for generic two-component fermion and scalar fields.
After spontaneous breaking of the electroweak symmetry, ($\langle H_1^1 \rangle=v_1/\sqrt{2} $ and $\langle H_2^2\rangle=v_2/\sqrt{2}$),
we have the following set of mass terms written in the 4-component spinor notation
so that
\beq
-{\cal L}_m= \bar\xi_R^T (M_f) \xi_L +\bar\eta_R^T(M_{\ell}) \eta_L +\text{H.c.},
\eeq
where the basis vectors in which the mass matrix is written is given by
\begin{gather}
\bar\xi_R^T= \left(\begin{matrix}\bar \nu_{\tau R} & \bar N_R & \bar \nu_{\mu R}
&\bar \nu_{e R} &\bar \nu_{4 R}\end{matrix}\right),\nonumber\\
\xi_L^T= \left(\begin{matrix} \nu_{\tau L} &  N_L &  \nu_{\mu L}
& \nu_{e L}&\nu_{4 L} \end{matrix}\right) \ ,\nonumber\\
\bar\eta_R^T= \left(\begin{matrix}\bar{\tau_ R} & \bar E_R & \bar{\mu_ R}
&\bar{e_ R}
&\bar{\ell}_{4R}
 \end{matrix}\right),\nonumber\\
\eta_L^T= \left(\begin{matrix} {\tau_ L} &  E_L &  {\mu_ L}
& {e_ L}
&\ell_{4L}
\end{matrix}\right) \ ,
\end{gather}
and the mass matrix $M_f$ of neutrinos  is given by

\beqn
M_f=
 \left(\begin{matrix} f'_1 v_2/\sqrt{2} & f_5 & 0 & 0 &0\cr
 -f_3 & f_2 v_1/\sqrt{2} & -f_3' & -f_3'' &-h_6\cr
0&f_5'&h_1' v_2/\sqrt{2} & 0 &0\cr
0 & f_5'' & 0 & h_2' v_2/\sqrt{2}&0\cr
0&h_8&0&0&  y_5' v_2/\sqrt{2}
\end{matrix} \right)\ .
\label{7}
\eeqn
We define the matrix element $(22)$ of the mass matrix as $m_N$ so that
\beqn
m_N= f_2 v_1/\sqrt 2.
\label{mn}
\eeqn
The mass matrix is not hermitian and thus one needs bi-unitary transformations to diagonalize it.
We define the bi-unitary transformation so that

\beq
D^{\nu \dagger}_R (M_f) D^\nu_L=\text{diag}(m_{\psi_1},m_{\psi_2},m_{\psi_3}, m_{\psi_4}, m_{\psi_5}  ).
\label{Dnu}
\eeq
where
$\psi_1, \psi_2, \psi_3, \psi_4, \psi_5$ are the mass eigenstates for the neutrinos.
In the limit of no mixing
we identify $\psi_1$ as the light tau neutrino, $\psi_2$ as the
heavier mass mirror eigen state,  $\psi_3$ as the muon neutrino, $\psi_4$ as the electron neutrino and $\psi_5$ as the other heavy 4-sequential generation neutrino.
A similar analysis goes to the lepton mass matrix $M_\ell$ where
\beqn
M_\ell=
 \left(\begin{matrix} f_1 v_1/\sqrt{2} & f_4 & 0 & 0 &0\cr
 f_3 & f'_2 v_2/\sqrt{2} & f_3' & f_3'' &h_6\cr
0&f_4'&h_1 v_1/\sqrt{2} & 0 &0\cr
0 & f_4'' & 0 & h_2 v_1/\sqrt{2}&0 \cr
0&h_7&0&0& y_5 v_1/\sqrt{2}
\end{matrix} \right)\ .
\label{7}
\eeqn
We introduce now the mass parameter $m_E$  for the (22) element of the mass matrix above so that
\beqn
m_E=  f_2' v_2/\sqrt 2.
\label{me}
\eeqn
CP phases that arise from the new sector are defined so that
\beqn
f_i= |f_i|e^{i\chi_i}, ~f'_i= |f'_i|e^{i\chi'_i}, ~f^{''}_i= |f^{''}_i|e^{i\chi^{''}_i}  ~~(i=3,4,5)\nonumber\\
h_k= |h_k|e^{i\chi_k}, ~~k=6,7,8\,.~~~~~~~~~~~~~~~~~~~~~~~~~~~~~~~~~~~~
\label{def-phases}
\eeqn

{
As in the neutrino mass matrix case, 
the charged slepton mass matrix is not hermitian and thus one needs again a bi-unitary transformations to diagonalize it.
We define the bi-unitary transformation so that

\beq
D^{\tau \dagger}_R (M_\ell) D^\tau_L=\text{diag}(m_{\tau_1},m_{\tau_2},m_{\tau_3}, m_{\tau_4}, m_{\tau_5}  ).
\label{Dtau}
\eeq
where
$\tau_\alpha$ ($\alpha=1-5$)  are the mass eigenstates for the charged lepton matrix.
}

The mass squared matrices of the slepton-mirror slepton and sneutrino-mirror sneutrino  sectors come from three sources: the F term, the D term of the potential and the soft SUSY breaking terms. After spontaneous breaking of the electroweak
symmetry the Lagrangian is given by
\beq
{\cal L}= {\cal L}_F +{\cal L}_D + {\cal L}_{\rm soft}\ ,
\eeq
where   $ {\cal L}_F$ is deduced from $-{\cal L}_F=F_i F^*_i$, while the ${\cal L}_D$ is given by
\begin{align}
-{\cal L}_D&=\frac{1}{2} m^2_Z \cos^2\theta_W \cos 2\beta \{\tilde \nu_{\tau L} \tilde \nu^*_{\tau L} -\tilde \tau_L \tilde \tau^*_L
+\tilde \nu_{\mu L} \tilde \nu^*_{\mu L} -\tilde \mu_L \tilde \mu^*_L
+\tilde \nu_{e L} \tilde \nu^*_{e L} -\tilde e_L \tilde e^*_L \nonumber \\
&+\tilde E_R \tilde E^*_R -\tilde N_R \tilde N^*_R
+\tilde \nu_{4 L} \tilde \nu^*_{4 L} -\tilde \ell_{4L} \tilde \ell^*_{4L}
\}
+\frac{1}{2} m^2_Z \sin^2\theta_W \cos 2\beta \{\tilde \nu_{\tau L} \tilde \nu^*_{\tau L}
 +\tilde \tau_L \tilde \tau^*_L
+\tilde \nu_{\mu L} \tilde \nu^*_{\mu L} +\tilde \mu_L \tilde \mu^*_L \nonumber \\
&+\tilde \nu_{e L} \tilde \nu^*_{e L} +\tilde e_L \tilde e^*_L
+\tilde \nu_{4 L} \tilde \nu^*_{4 L}
 +\tilde \ell_{4L} \tilde \ell^*_{4L}\nonumber\\
&-\tilde E_R \tilde E^*_R -\tilde N_R \tilde N^*_R +2 \tilde E_L \tilde E^*_L -2 \tilde \tau_R \tilde \tau^*_R
-2 \tilde \mu_R \tilde \mu^*_R -2 \tilde e_R \tilde e^*_R
-2 \tilde \ell_{4 R} \tilde \ell^*_{4 R}
\}.
\label{12}
\end{align}

For ${\cal L}_{\rm soft}$ we assume the following form
\begin{align}
-{\cal L}_{\text{soft}}&=\tilde M^2_{\tau L} \tilde \psi^{i*}_{\tau L} \tilde \psi^i_{\tau L}
+\tilde M^2_{\chi} \tilde \chi^{ci*} \tilde \chi^{ci}
+\tilde M^2_{\mu L} \tilde \psi^{i*}_{\mu L} \tilde \psi^i_{\mu L}\nonumber\\
&+\tilde M^2_{e L} \tilde \psi^{i*}_{e L} \tilde \psi^i_{e L}
+\tilde M^2_{\nu_\tau} \tilde \nu^{c*}_{\tau L} \tilde \nu^c_{\tau L}
 +\tilde M^2_{\nu_\mu} \tilde \nu^{c*}_{\mu L} \tilde \nu^c_{\mu L} \nonumber \\
&
+\tilde M^2_{4 L} \tilde \psi^{i*}_{4 L} \tilde \psi^i_{4 L}
+\tilde M^2_{\nu_4} \tilde \nu^{c*}_{4 L} \tilde \nu^c_{4 L}
+\tilde M^2_{\nu_e} \tilde \nu^{c*}_{e L} \tilde \nu^c_{e L}
+\tilde M^2_{\tau} \tilde \tau^{c*}_L \tilde \tau^c_L
+\tilde M^2_{\mu} \tilde \mu^{c*}_L \tilde \mu^c_L\nonumber\\
&
+\tilde M^2_{e} \tilde e^{c*}_L \tilde e^c_L
+\tilde M^2_E \tilde E^*_L \tilde E_L
 + \tilde M^2_N \tilde N^*_L \tilde N_L
+\tilde M^2_{4} \tilde \ell^{c*}_{4L} \tilde \ell^c_{4L}
 \nonumber \\
&+\epsilon_{ij} \{f_1 A_{\tau} H^i_1 \tilde \psi^j_{\tau L} \tilde \tau^c_L
-f'_1 A_{\nu_\tau} H^i_2 \tilde \psi ^j_{\tau L} \tilde \nu^c_{\tau L}
+h_1 A_{\mu} H^i_1 \tilde \psi^j_{\mu L} \tilde \mu^c_L
-h'_1 A_{\nu_\mu} H^i_2 \tilde \psi ^j_{\mu L} \tilde \nu^c_{\mu L} \nonumber \\
&+h_2 A_{e} H^i_1 \tilde \psi^j_{e L} \tilde e^c_L
-h'_2 A_{\nu_e} H^i_2 \tilde \psi ^j_{e L} \tilde \nu^c_{e L}
+f_2 A_N H^i_1 \tilde \chi^{cj} \tilde N_L
-f'_2 A_E H^i_2 \tilde \chi^{cj} \tilde E_L \nonumber\\
&
+y_5 A_{4\ell} H^i_1 \tilde \psi^j_{4 L} \tilde \ell^c_{4L}
-y'_5 A_{4\nu} H^i_2 \tilde \psi ^j_{4 L} \tilde \nu^c_{4 L}
+\text{H.c.}\}\ .
\label{13}
\end{align}

We define the scalar mass squared   matrix $M^2_{\tilde \tau}$  in the basis
\beq
 (\tilde  \tau_L, \tilde E_L, \tilde \tau_R,
\tilde E_R, \tilde \mu_L, \tilde \mu_R, \tilde e_L, \tilde e_R, \tilde\ell_{4L}, \tilde\ell_{4R}).
\eeq
 We  label the matrix  elements of these as $(M^2_{\tilde \tau})_{ij}= M^2_{ij}$ where the elements of the matrix are given by
\begin{gather}
M^2_{11}=\tilde M^2_{\tau L} +\frac{v^2_1|f_1|^2}{2} +|f_3|^2 -m^2_Z \cos 2 \beta \left(\frac{1}{2}-\sin^2\theta_W\right), \nonumber\\
M^2_{22}=\tilde M^2_E +\frac{v^2_2|f'_2|^2}{2}+|f_4|^2 +|f'_4|^2+|f''_4|^2
+|h_7|^2
 +m^2_Z \cos 2 \beta \sin^2\theta_W, \nonumber\\
M^2_{33}=\tilde M^2_{\tau} +\frac{v^2_1|f_1|^2}{2} +|f_4|^2 -m^2_Z \cos 2 \beta \sin^2\theta_W, \nonumber\\
M^2_{44}=\tilde M^2_{\chi} +\frac{v^2_2|f'_2|^2}{2} +|f_3|^2 +|f'_3|^2+|f''_3|^2
+|h_6|^2
+m^2_Z \cos 2 \beta \left(\frac{1}{2}-\sin^2\theta_W\right), \nonumber\\
M^2_{55}=\tilde M^2_{\mu L} +\frac{v^2_1|h_1|^2}{2} +|f'_3|^2 -m^2_Z \cos 2 \beta \left(\frac{1}{2}-\sin^2\theta_W\right), \nonumber\\
M^2_{66}=\tilde M^2_{\mu} +\frac{v^2_1|h_1|^2}{2}+|f'_4|^2 -m^2_Z \cos 2 \beta \sin^2\theta_W, \nonumber
\end{gather}

\begin{gather}
M^2_{77}=\tilde M^2_{e L} +\frac{v^2_1|h_2|^2}{2}+|f''_3|^2 -m^2_Z \cos 2 \beta \left(\frac{1}{2}-\sin^2\theta_W\right), \nonumber\\
M^2_{88}=\tilde M^2_{e} +\frac{v^2_1|h_2|^2}{2}+|f''_4|^2 -m^2_Z \cos 2 \beta \sin^2\theta_W, \nonumber\\
M^2_{99}=\tilde M^2_{4 L} +\frac{v^2_1|y_5|^2}{2} +|h_6|^2 -m^2_Z \cos 2 \beta \left(\frac{1}{2}-\sin^2\theta_W\right), \nonumber\\
M^2_{10 10}=\tilde M^2_{4} +\frac{v^2_1|y_5|^2}{2} +|h_7|^2 -m^2_Z \cos 2 \beta \sin^2\theta_W, \nonumber\\
M^2_{12}=M^{2*}_{21}=\frac{ v_2 f'_2f^*_3}{\sqrt{2}} +\frac{ v_1 f_4 f^*_1}{\sqrt{2}} ,\nonumber\\
M^2_{13}=M^{2*}_{31}=\frac{f^*_1}{\sqrt{2}}(v_1 A^*_{\tau} -\mu v_2),\nonumber\\
M^2_{14}=M^{2*}_{41}=0, M^2_{15} =M^{2*}_{51}=f'_3 f^*_3,\nonumber\\
 M^{2}_{16}= M^{2*}_{61}=0,  M^{2}_{17}= M^{2*}_{71}=f''_3 f^*_3,  M^{2}_{18}= M^{2*}_{81}=0,
M^2_{23}=M^{2*}_{32}=0,\nonumber\\
M^2_{24}=M^{2*}_{42}=\frac{f'^*_2}{\sqrt{2}}(v_2 A^*_{E} -\mu v_1),  M^2_{25} = M^{2*}_{52}= \frac{ v_2 f'_3f'^*_2}{\sqrt{2}} +\frac{ v_1 h_1 f^*_4}{\sqrt{2}} ,\nonumber\\
 M^2_{26} =M^{2*}_{62}=0,  M^2_{27} =M^{2*}_{72}=  \frac{ v_2 f''_3f'^*_2}{\sqrt{2}} +\frac{ v_1 h_1 f'^*_4}{\sqrt{2}},  M^2_{28} =M^{2*}_{82}=0, \nonumber\\
M^2_{34}=M^{2*}_{43}= \frac{ v_2 f_4 f'^*_2}{\sqrt{2}} +\frac{ v_1 f_1 f^*_3}{\sqrt{2}}, M^2_{35} =M^{2*}_{53} =0, M^2_{36} =M^{2*}_{63}=f_4 f'^*_4,\nonumber\\
 M^2_{37} =M^{2*}_{73} =0,  M^2_{38} =M^{2*}_{83} =f_4 f''^*_4,
M^2_{45}=M^{2*}_{54}=0, M^2_{46}=M^{2*}_{64}=\frac{ v_2 f'_2 f'^*_4}{\sqrt{2}} +\frac{ v_1 f'_3 h^*_1}{\sqrt{2}}, \nonumber\\
 M^2_{47} =M^{2*}_{74}=0,  M^2_{48} =M^{2*}_{84}=  \frac{ v_2 f'_2f''^*_4}{\sqrt{2}} +\frac{ v_1 f''_3 h^*_2}{\sqrt{2}},\nonumber\\
M^2_{56}=M^{2*}_{65}=\frac{h^*_1}{\sqrt{2}}(v_1 A^*_{\mu} -\mu v_2),
 M^2_{57} =M^{2*}_{75}=f''_3 f'^*_3,  M^2_{58} =M^{2*}_{85}=0,  M^2_{67} =M^{2*}_{76}=0,\nonumber\\
 M^2_{68} =M^{2*}_{86}=f'_4 f''^*_4,  M^2_{78}=M^{2*}_{87}=\frac{h^*_2}{\sqrt{2}}(v_1 A^*_{e} -\mu v_2)\nonumber\\
M^{2}_{19}= M^{2*}_{91}=f^*_3 h_6,  M^{2}_{1 10}= M^{2*}_{10 1}=0,\nonumber\\
 M^2_{29} =M^{2*}_{92}=\frac{ v_1 y_5h^*_7}{\sqrt{2}} +\frac{ v_2 h_6 f'^*_2}{\sqrt{2}},  M^2_{2 10} =M^{2*}_{10 2}=0,\nonumber\\
 M^{2}_{39}= M^{2*}_{93}=0,  M^{2}_{3 10}= M^{2*}_{10 3}=f_4 h^*_7,\nonumber\\
 M^2_{49} =M^{2*}_{94}=0,  M^2_{4 10} =M^{2*}_{10 4}=\frac{ v_2 f'_2h^*_7}{\sqrt{2}} +\frac{ v_1 h_6 y^*_5}{\sqrt{2}},\nonumber\\
M^{2}_{59}= M^{2*}_{95}=f'^*_3 h_6,  M^{2}_{5 10}= M^{2*}_{10 5}=0,\nonumber\\
M^{2}_{69}= M^{2*}_{96}=0,  M^{2}_{6 10}= M^{2*}_{10 6}=f'_4 h^*_7,\nonumber\\
M^{2}_{79}= M^{2*}_{97}=f''^*_3 h_6,  M^{2}_{7 10}= M^{2*}_{10 7}=0,\nonumber\\
M^{2}_{89}= M^{2*}_{98}=0,  M^{2}_{8 10}= M^{2*}_{10 8}=f''_5 h^*_7,\nonumber\\
M^2_{9 10}=M^{2*}_{10 9}=\frac{y^*_5}{\sqrt{2}}(v_1 A^*_{4\ell} -\mu v_2)
\label{14}
\end{gather}
   {We assume that  the masses that enter the mass squared matrix for the scalars are all of electroweak size.
   This mass squared matrix is hermitian and can be diagonalized with a  unitary transformation.} 

\beq
 \tilde D^{\tau \dagger} M^2_{\tilde \tau} \tilde D^{\tau} = diag (M^2_{\tilde \tau_1},
M^2_{\tilde \tau_2}, M^2_{\tilde \tau_3},  M^2_{\tilde \tau_4},  M^2_{\tilde \tau_5},  M^2_{\tilde \tau_6},  M^2_{\tilde \tau_7},  M^2_{\tilde \tau_8} M^2_{\tilde \tau_9},  M^2_{\tilde \tau_{10}} )
\label{Dtildetau}
\eeq

The  mass squared  matrix in the sneutrino sector has a similar structure. In the basis
\beq
(\tilde  \nu_{\tau L}, \tilde N_L,
 \tilde \nu_{\tau R}, \tilde N_R, \tilde  \nu_{\mu L},\tilde \nu_{\mu R}, \tilde \nu_{e L}, \tilde \nu_{e R},
\tilde \nu_{4 L}, \tilde \nu_{4 R}
 )
\eeq
 {the sneutrino mass squared matrix
$(M^2_{\tilde\nu})_{ij}=m^2_{ij}$ has elements given by}
\begin{gather}
m^2_{11}=\tilde M^2_{\tau L} +\frac{v^2_2}{2}|f'_1|^2 +|f_3|^2 +\frac{1}{2}m^2_Z \cos 2 \beta,  \nonumber\\
m^2_{22}=\tilde M^2_N +\frac{v^2_1}{2}|f_2|^2 +|f_5|^2 +|f'_5|^2+|f''_5|^2+|h_8|^2, \nonumber\\
m^2_{33}=\tilde M^2_{\nu_\tau} +\frac{v^2_2}{2}|f'_1|^2 +|f_5|^2,  \nonumber\\
m^2_{44}=\tilde M^2_{\chi} +\frac{v^2_1}{2}|f_2|^2 +|f_3|^2 +|f'_3|^2+|f''_3|^2+|h_6|^2 -\frac{1}{2}m^2_Z \cos 2 \beta, \nonumber\\
m^2_{55}=\tilde M^2_{\mu L} +\frac{v^2_2}{2}|h'_1|^2 +|f'_3|^2 +\frac{1}{2}m^2_Z \cos 2 \beta,  \nonumber\\
m^2_{66}=\tilde M^2_{\nu_\mu} +\frac{v^2_2}{2}|h'_1|^2 +|f'_5|^2,  \nonumber\\
m^2_{77}=\tilde M^2_{e L} +\frac{v^2_2}{2}|h'_2|^2+|f''_3|^2+\frac{1}{2}m^2_Z \cos 2 \beta,  \nonumber\\
m^2_{88}=\tilde M^2_{\nu_e} +\frac{v^2_2}{2}|h'_2|^2 +|f''_5|^2,  \nonumber\\
m^2_{99}=\tilde M^2_{4 L} +\frac{v^2_2}{2}|y'_5|^2 +|h_6|^2+\frac{1}{2}m^2_Z \cos 2 \beta,  \nonumber\\
m^2_{10 10}=\tilde M^2_{\nu 4} +|h_8|^2 +\frac{v^2_2}{2}|y'_5|^2 , \nonumber\\
m^2_{12}=m^{2*}_{21}=\frac{v_2 f_5 f'^*_1}{\sqrt{2}}-\frac{ v_1 f_2 f^*_3}{\sqrt{2}},\nonumber\\
m^2_{13}=m^{2*}_{31}=\frac{f'^*_1}{\sqrt{2}}(v_2 A^*_{\nu_\tau} -\mu v_1),
m^2_{14}=m^{2*}_{41}=0,\nonumber\\
m^2_{15}=m^{2*}_{51}= f'_3 f^*_3, m^2_{16}=m^{2*}_{61}=0,\nonumber\\
m^2_{17}=m^{2*}_{71}= f''_3 f^*_3, m^2_{18}=m^{2*}_{81}=0,\nonumber\\
m^2_{23}=m^{2*}_{32}=0,
m^2_{24}=m^{2*}_{42}=\frac{f^*_2}{\sqrt{2}}(v_{1}A^*_N-\mu v_2), m^2_{25}=m^{2*}_{52}=-\frac{v_{1}f^*_2 f'_3}{\sqrt{2}}+\frac{h'_1 v_2 f'^*_5}{\sqrt{2}},\nonumber\\
m^2_{26}=m^{2*}_{62}=0, m^2_{27}=m^{2*}_{72}=-\frac{v_{1}f^*_2 f''_3}{\sqrt{2}}+\frac{h'_2 v_2 f''^*_5}{\sqrt{2}},
\end{gather}

\begin{gather}
m^2_{28}=m^{2*}_{82}=0, m^2_{34}=m^{2*}_{43}=\frac{v_1 f^*_2 f_5}{\sqrt{2}}-\frac{v_2 f'_1 f^*_3}{\sqrt{2}},\nonumber\\
m^2_{35}=m^{2*}_{53}=0, m^2_{36}=m^{2*}_{63}=f_5 f'^*_5, m^2_{37}=m^{2*}_{73}=0, m^2_{38}=m^{2*}_{83}=f_5 f''^*_5, m^2_{45}=m^{2*}_{54}=0, \nonumber\\
m^2_{46}=m^{2*}_{64}=-\frac{h'^*_1 v_2 f'_3}{\sqrt{2}}+\frac{v_1 f_2 f'^*_5}{\sqrt{2}}, m^2_{47}=m^{2*}_{74}=0, \nonumber\\
m^2_{48}=m^{2*}_{84}=\frac{v_1 f_2 f''^*_5}{\sqrt{2}}-\frac{v_2 h'^*_2 f''_3}{\sqrt{2}}, m^2_{56}=m^{2*}_{65}=\frac{h'^*_1}{\sqrt{2}}(v_2 A^*_{\nu_\mu}-\mu v_1), \nonumber\\
m^2_{57}=m^{2*}_{75}= f''_3 f'^*_3, m^2_{58}=m^{2*}_{85}=0, m^2_{67}=m^{2*}_{76}=0, \nonumber\\
m^2_{68}=m^{2*}_{86}= f'_5 f''^*_5, m^2_{78}=m^{2*}_{87}=\frac{h'^*_2}{\sqrt{2}}(v_2 A^*_{\nu_e}-\mu v_1),\nonumber\\
m^2_{19}=m^{2*}_{91}= h_6 f^*_3, m^2_{1 10}=m^{2*}_{10 1}=0,\nonumber\\
m^2_{29}=m^{2*}_{92}=-\frac{f_2 v_1 h_6}{\sqrt{2}}+\frac{v_2 h_8 y^*_5}{\sqrt{2}}, m^2_{2 10}=m^{2*}_{10 2}=0, \nonumber\\
m^2_{39}=m^{2*}_{93}=0, m^2_{3 10}=m^{2*}_{10 3}=f_5 h^*_8,\nonumber\\
m^2_{49}=m^{2*}_{94}=0,
m^2_{4 10}=m^{2*}_{10 4}=-\frac{v_2 y'_5 h_6}{\sqrt{2}}+\frac{v_1 h^*_8 f_2}{\sqrt{2}},\nonumber\\
m^2_{59}=m^{2*}_{95}= h_6 f'^*_3, m^2_{5 10}=m^{2*}_{10 5}=0,\nonumber\\
m^2_{69}=m^{2*}_{96}=0, m^2_{6 10}=m^{2*}_{10 6}=f'_5 h^*_8,\nonumber\\
m^2_{79}=m^{2*}_{97}= h_6 f''^*_3, m^2_{7 10}=m^{2*}_{10 7}=0,\nonumber\\
m^2_{89}=m^{2*}_{98}=0, m^2_{8 10}=m^{2*}_{10 8}=f''_5 h^*_8,\nonumber\\
m^2_{9 10}=m^{2*}_{10 9}=\frac{y'_5}{\sqrt{2}}(v_2 A^*_{4 \nu}-\mu v_1).
\label{15}
\end{gather}
  Again as in the charged lepton sector we assume that all the masses are of the electroweak size so all the terms enter in the mass squared matrix.  This mass squared  matrix can be diagonalized  by the unitary transformation

  \beq
\tilde D^{\nu\dagger} M^2_{\tilde \nu} \tilde D^{\nu} = \text{diag} (M^2_{\tilde \nu_1}, M^2_{\tilde \nu_2}, M^2_{\tilde \nu_3},  M^2_{\tilde \nu_4},M^2_{\tilde \nu_5},  M^2_{\tilde \nu_6}, M^2_{\tilde \nu_7}, M^2_{\tilde \nu_8},
 M^2_{\tilde \nu_9},  M^2_{\tilde \nu_{10}}
 ).
 \label{Dtildenu}
\eeq

\end{document}